\documentclass{emulateapj}

\newcommand{\msolar}{M$_\odot$}
\def\la{\mathrel{\mathpalette\fun <}}

\def\fun#1#2{\lower3.6pt\vbox{\baselineskip0pt\lineskip.9pt
         \ialign{$\mathsurround=0pt#1\hfill##\hfil$\crcr#2\crcr\sim 
\crcr}}}
\def\be#1{\begin{equation}\label{eq:#1}}
\def\ee{\end{equation}}

\def\bea#1{\begin{eqnarray}\label{eq:#1}}
\def\ee{\end{equation}}
\def\eea{\end{eqnarray}}
\def\la{\mathrel{\mathpalette\fun <}}

\def\kmskpc{{\rm ~km~s^{-1}~kpc^{-1}}}

\begin{document}
\submitted{Submitted to ApJ \today}

\title{ANATOMY OF THE BAR INSTABILITY IN CUSPY DARK MATTER HALOS}

\author{
John Dubinski\altaffilmark{1},
Ingo Berentzen\altaffilmark{2} and
Isaac Shlosman\altaffilmark{3,4}
}

\altaffiltext{1}{Department of Astronomy and Astrophysics,
University of Toronto, 50 St. George Street, Toronto, ON M5S 3H4,  
Canada;
dubinski@astro.utoronto.ca}
\altaffiltext{2}{
Astronomisches Rechen-Institut, Zentrum f\"ur Astronomie, 
Universit\"at Heidelberg,
M\"onchhofstr. 12-14 69120,
Heidelberg, Germany;
iberent@ari.uni-heidelberg.de}
\altaffiltext{3}{JILA, University of Colorado, Boulder, CO 80309-0440,
USA; shlosman@pa.uky.edu}
\altaffiltext{4}{Department of Physics and Astronomy, University of
Kentucky, Lexington, KY 40506-0055, USA}

\begin{abstract}
We examine the bar instability in galactic models with an exponential disk
and a cuspy dark matter (DM) halo with a Navarro-Frenk-White (NFW) 
cosmological density profile.
The equilibrium models are constructed from a 3-integral composite distribution function
but subject to the bar instability.
We generate a sequence of models with a range of mass resolution
from 1.8K to 18M particles in the disk and 10K to 100M particles in 
the halo along with a multi-mass model with an effective
resolution of $\sim 10^{10}$ particles.   We describe
how mass resolution affects the bar instability,
including its linear growth phase, the buckling instability, pattern
speed decay through the resonant transfer of angular momentum to the  
DM halo, and the possible destruction of the halo cusp.
Our higher resolution simulations show a converging spectrum of 
discrete resonance interactions between the bar and DM halo orbits.  
As the pattern speed decays, orbital resonances sweep through most of 
the DM halo phase space and widely distribute angular momentum among 
the halo particles. The halo does not develop a flat density core 
and preserves the cusp, except in the region dominated by 
gravitational softening.  The formation of the bar increases 
the central stellar density and the DM is compressed adiabatically
increasing the halo central density by $1.7\times$. 
Overall, the evolution of the bar displays a convergent behavior for  
halo particle numbers between 1M and 10M particles, when comparing 
bar growth, pattern speed evolution, the DM halo density profile 
and a nonlinear analysis of the orbital resonances. 
Higher resolution simulations clearly illustrate 
the importance of discrete resonances in transporting the angular 
momentum from the bar to the halo.
\end{abstract}



\section{Introduction}

More than 2/3 of disk galaxies host stellar bars
\citep[e.g.,][]{kna00,gro04,mar07} and evolution of this fraction with
redshift is a matter of an ongoing debate \citep[e.g.,][]{jog04,she08}.
Numerical simulations of disk galaxies have shown that bars
form either as a result of a global gravitational instability 
\citep[e.g.,][]{too81,sel93} or they are triggered by galaxy interactions 
\citep[e.g.,][]{byr86,nog87} and
interactions with DM substructure 
\citep[e.g.,][]{gau06,dub08a,rom08}.
A large body of theoretical work on the bar instability
has examined the properties of bars that emerge in initially  
unstable disks in $N$-body simulations.  While these experiments explore  
an idealized picture of bar formation, they reveal important aspects 
of the phenomenology of the bar instability, including bar growth 
within the corotation (CR) radius, the vertical buckling instability, and the
transport of angular momentum through gravitational torques from  
resonant orbits in the outer disk and the surrounding dark matter (DM) halo.
The importance of the resonance nature of angular momentum loss by  
bars and spirals was first pointed out by \citet{lyn72}. Angular
momentum transfer was studied subsequently both in idealized models 
with rigid bars in live halos \citep{wei85,her92,wei02,wei07a}; 
and self-consistent $N$-body simulations with bar-unstable disks 
\citep[e.g.,][]{sel80,ath96,deb98,val03,one03}, 
with resonant transfer mechanisms being explored explicitly in some 
studies \citep[e.g.][]{ath02,hol05,mar06,cev07}. 
These studies have shown that the halo absorbs angular momentum from the bar 
that leads to the decline of the bar pattern speed.

Previous results reveal a close connection between numerical bars and
observed galactic systems in many structural details, including a link between
the peanut-shaped bulges and (buckled) bars
\citep[e.g.,][]{com81,com90,rah91,ber98,pat02,mar06,deb06}.
The observational determination of bar pattern speeds 
\citep[e.g.,][]{ken87,mer95,cor07} suggest that stellar bars are 
predominantly ``fast" (but see \citet{rau08} for a different view) meaning that
they are near the maximum possible length of the CR radius
permitted by the orbital dynamics \citep{con80,ath92}. If evolved for too long, 
the numerical bars can appear ``slow" with lengths significantly shorter 
than the CR  radius and pattern speeds that seem abnormally low when 
compared to observations of real barred galaxies \citep{deb98,deb00}. 
However, at higher resolution, even collisionless numerical bars seem 
to grow in length towards their CR radius by capturing disk orbits 
and so remain ``fast" \citep{mar06}. 
Furthermore, the addition of gas
may stabilize the bar against braking and results in its speedup instead for
prolonged time periods \citep{rom08}.

Some studies claim that bars may destroy 
the cuspy profiles of DM halos predicted by the CDM
cosmology \citep[e.g.,][]{dub91,nav96}, thus alleviating an apparent  
contradiction between the inferred density profiles of DM halos from 
galactic rotation curves and this theoretical expectation in some cases 
\citep{wei02,hol05,wei07b}.  Simulations demonstrating cusp destruction 
use rigid, ellipsoidal bars --- their applicability to self-consistent 
dynamical systems is suspect.  Also, there has been some concern about
artifical $m=1$ instabilities arising from using a fixed center in
$N$-body field expansion methods \citep{sel03,mcm05}.  
Current studies have obtained
contradictory results on the efficiency of angular momentum transport
to the cusp.  \citet{wei07a} have emphasized the importance of 
numerical resolution and, specifically, of the total particle number 
in simulations.  Since the transport of angular momentum operates mainly 
through 
low order resonances between the bar pattern speed and halo orbital 
frequencies, only a small fraction of the halo mass participates.  
Without adequate particle numbers then, they argue that torques associated with 
resonant populations may be under-sampled, leading to a spurious calculation 
of angular momentum transport and, therefore, the evolution of the bar overall.
\citet{wei07b} estimate that at least $10^8$ particles and maybe more may be necessary 
to sample the phase-space densely enough to converge to the correct answer.
Recently, Sellwood (2008) has disputed this claim in simulations with rigid bars
in spherical, isotropic halos with $\sim 10^8$ particles arguing that the resonances
are broader than they claim

In this paper, we address the issue of the numerical convergence of bar 
evolution using a series of $N$-body simulations of the bar instability in a 
self-consistent model galaxy. We analyze bar growth in a
bar-unstable $N$-body disk. In contrast to other work, we employ new 
galactic models based on the methods of \citet{wid05}, and carry out 
simulations with substantially greater numerical resolution 
than reported in the literature.  The galaxy is described 
by a well-defined distribution function for an exponential disk embedded 
within a DM halo with an $r^{-1}$ density cusp, based on a truncated 
Navarro, Frenk \& White (1996, NFW) profile.
These models are formally in 
dynamical equilibrium but are bar-unstable.  Since they are defined by a 
distribution function, their $N$-body realizations are equivalent, independent
of the particle numbers. Hence, this study can probe the effect of numerical 
resolution on collisionless galaxy evolution.  
Our goal is to quantify the behavior of a 
number of specific parameters describing the bar instability as a function 
of particle number, including the bar strength amplitude, 
$A_2$, as given by the $m=2$ Fourier mode, as well as its pattern 
speed evolution, 
angular momentum transport, and evolution of the DM density profiles, 
particularly in the region within the halo characteristic NFW scale 
radius, $r_s$.  We also perform an orbital spectral analysis of halo and 
disk particles, to quantify the effect of the low order resonances 
responsible for angular momentum transport \citep{ath02,mar06}.

The plan of the paper is as follows.  In \S 2, we provide a description of
the galactic models and the $N$-body experiments to study the bar instability.
In \S 3, we present results on the bar growth and the evolution of pattern 
speed as a function of numerical resolution.  In \S 4, we examine the 
evolution the DM halo density profile as a function of numerical resolution.
In \S 5, we study the low order resonances between the bar and the
halo particles using orbital integrations and spectral analysis and 
again compare results at different resolutions.  We also examine the 
details of the evolution of the halo phase space density in our highest 
resolution models.  We conclude with a discussion of the 
importance of numerical resolution in these experiments and comment on 
the reliability of current work in studies of disk galaxy formation and 
dynamics.

\section{Methods}
\subsection{Initial conditions: An exponential disk with a cuspy dark halo}

The main goal of this study is to characterize the  bar instability in terms
of mass resolution.  The galaxy models of \citet{wid05} (WD models  
herein) are ideal for this purpose since they are derived from a composite
3-integral distribution function (DF) $f \equiv f_{disk}(E,L_z,E_z) +
f_{halo}(E)$. The disk model has an exponential radial profile and 
${\rm sech}^2 z$ vertical profile.  The disk DF $f_{disk}$ is a 3D 
extension of the 2D function introduced by \citet{shu69} using the vertical 
energy $E_z = 1/2 \dot{z}^2 + \Phi(R,z) - \Phi(R,z=0)$ as an approximate 
third integral \citep{kui95}.  This DF applies in the epicyclic approximation 
with $\sigma_{R, \phi,z} \ll v_c$ and so the vertical energy is 
approximately constant.  This leads to triaxial velocity ellipsoids in 
the disk models as seen in real spiral galaxies.  These models generally 
provide 
near equilibrium initial conditions and show negligible transient behavior at
startup \citep{wid05}.  The halo DF $f_{halo}$ describes a truncated 
spherical, isotropic NFW model.  When the two DFs are combined, the net 
halo density profile changes slightly from the NFW form and is flattened 
along the $z$-axis near the center, but preserves the $r^{-1}$ central cusp.  
A suitable choice of parameters allows the construction of a realistic 
model of bulgeless spiral galaxy with a cosmologically inspired DM halo.
Since the models are derived from a distribution function, particle
distributions for $N$-body experiments can be generated by direct 
Monte-Carlo sampling.  

For the experiments described below, we
initially generate a model containing 18M disk particles and 100M
halo particles with both disk and halo particles having approximately 
the same mass.  The halo is non-rotating.  Lower resolution models are 
generated by subsampling this larger model in factors of ten and hence 
creating a sequence of models containing numbers of particles in the 
range $1.18\times 10^{4-8}$.  One further model is generated with 
a multi-mass DM halo to increase the 
particle number density in the core by another two orders of magnitude.
The particle mass is weighted as an approximate step function in angular 
momentum $m \sim m(L)$ such that low angular momentum particles near the halo
center below a characteristic angular momentum $L_c$ would have a lower mass.  
The number density at the center of this model is more than
$100\times$ greater so the effective particle number is $\approx 10^{10}$
for this simulation.  We describe the details for generating the multi-mass 
model below.  Our highest resolution simulations have large enough particle 
numbers to probe the divergence in numerical behavior discussed by 
\citep{wei07a}.

Each model is generated and simulated in units with $G=1$ and physical
quantities are of order unity.  We have designed the model as a proxy
for the Milky Way without a bulge, so natural units for this comparison
are $L=10$~kpc, $M=10^{11}$~\msolar, $V=207.8$~km~s$^{-1}$ and $T=47.2$~Myr.
By design, the model mass profile closely resembles
the one examined by \citet{mar06}.
Moreover, the central density cusp is better resolved and the initial
conditions are in a better equilibrium, since they are sampled from a  
DF.  Throughout this paper we present results in physical units. 

\begin{figure}
\begin{center}
\plotone{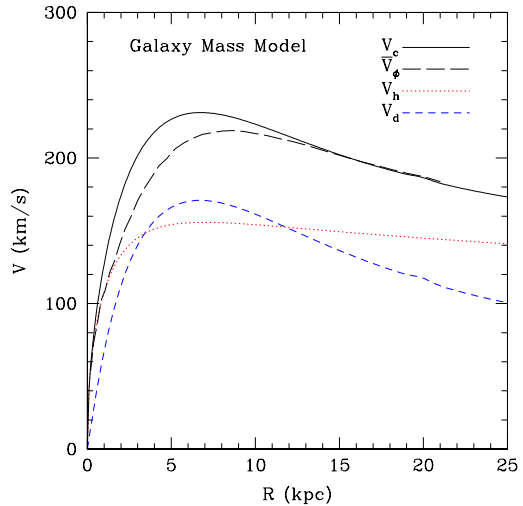}
\figcaption{
Initial circular velocity curve of the mass model showing the contributions
from the disk and the DM halo.  We also plot the mean tangential velocity 
in the disk to show the effect of an asymmetric drift on the rotation curve.
\label{fig-vrot1}
}
\end{center}
\end{figure}

\begin{figure}
\begin{center}
\plotone{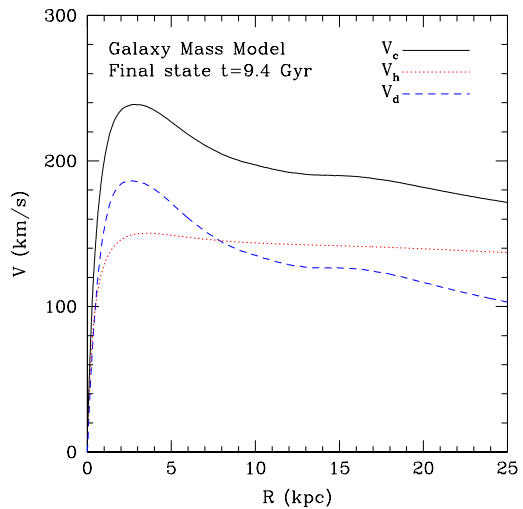}
\figcaption{
Final circular velocity curve of the barred galaxy mass model at $t=9.4$ Gyr.
We show the contributions from the barred disk and the DM halo.  The disk component 
is estimated by axisymmetrizing the barred disk and calculating $v_d^2=R\partial \Phi/\partial
R$.
\label{fig-vrot-final}
}
\end{center}
\end{figure}

The galaxy mass model is presented in Figure \ref{fig-vrot1} as a
rotation curve decomposition.  We use an exponential disk with
radial scale-length 2.85~kpc and an exponential vertical
scalelength of 250~pc and a total mass $5.5
\times 10^{10}$ \msolar.  The disk is truncated smoothly at $R=21$~kpc  
equivalent to 7.4 scale lengths. The NFW halo scale radius 
in the DF of the WD model is set to $r_s=10$~kpc but results in an
effective scale radius of $r_s=4.3$~kpc as measured by a 
least-squares fit to the density profile. The peak circular velocity of 
the DM halo is $v_{max}=0.77$ (160~km~s$^{-1}$).  We note that the smaller
scalelength is not due to an adiabatic contraction, but is the  
result of combining two distribution functions \citep{wid05} --- the extra
concentration of mass from the potential of the disk causes the halo
potential derived from the NFW DF to be more concentrated as well when
calculating the self-consistent potential for the model.  
The halo extends 
to a truncation radius of $r=260$~kpc and has a total mass 
$M=3.0$ units ($3.0 \times 10^{11}$ \msolar).  The final model
is a realistic facsimile of an exponential disk galaxy with a cuspy DM halo.
The square of the radial velocity dispersion $\sigma_R^2$ of these models 
follows the same exponential radial decline as the surface density with 
$\sigma_R^2 \sim \exp(-R/R_d)$.  We choose a central value
$\sigma_{R,0} = 104$~km~s$^{-1}$, so that the Toomre Q 
is $Q=1.1$ at $R=10$~kpc.  The disk is, therefore, relatively cold 
and responsive.  This model is in dynamical equilibrium but also is
strongly bar-unstable.  Our analysis focuses on the development of the bar
instability in simulations of this model with different particle numbers.
At this point, we also present the final state of the mass model after 9.4 Gyrs
of dynamical evolution for direct comparison to the initial state but defer the discussion
until later (Fig.~\ref{fig-vrot-final}).

\subsection{Multi-mass model}

A common way of increasing mass resolution with a number of particles is to use
a range of masses, assigning low mass particles to the center
where the action is and high mass particles to the periphery \citep{sig95}.
We therefore build an additional model that weights the halo particle mass as a 
monotonically increasing function of orbital angular momentum 
$L=|{\bf r}\times {\bf v}|$, to increase the number density of
particles in the region where the bar forms and where the low order
resonances occur.  The strategy is to define
a mass weighting function $W(L)$ 
such that particles with low angular
momentum and orbits with small pericenters also have small
mass, while those with large angular momentum and pericenters beyond
the edge of the disk have a higher mass.  The halo DF is normalized
by this weighting function, so that the number density of particles derived
from Monte Carlo sampling will be larger for smaller values of $L$.
In this way, the probability of selecting a particle with smaller $L$ is
greater than with larger $L$.  The biased number density is then corrected 
to represent the model with the original DF by multiplying the particle 
mass by $W(L)$.  The weighting function is normalized so that the mass of 
a particle in the initial distribution is given by
\begin{equation}
m_i = \frac{M_{halo}W(L_i)}{\sum_{i} W(L_i)}
\end{equation}

The choice of the functional form of $W(L)$ is arbitrary at some level
according to the needs of the problem but in our case it should be monotonically
increasing with $L$.  We use the step-like weighting function in $L$
\begin{equation}
W(L) = 1.0 + \frac{W_1-1}{1 + (L/L_{c})^{-\alpha}}
\end{equation}
where $L_c$ is a characteristic angular momentum for the step,
$\alpha>0$ is an exponent and $W_1$ is the asymptotic value of weighting
function for large angular momentum.  When $W$ is plotted versus $\log L$ 
it takes the form of a step function where the steepness of the transition
at $\log L_c$ depends on the choice of $\alpha$.  In practice, we truncate 
the function at minimum and maximum values of $L$ at $L_{min}$ and $L_{max}$ 
and  set the weight to the value at these limits beyond the endpoints.

After some experimentation, our final choices for these parameters are 
$W_1=10^4$, $L_{min}=10^{-3}$, $L_{c}=3$, $L_{max}=7$, and $\alpha=0.9$.
The choice of $L_c$ corresponds to particles moving at the circular velocity 
at a radius of $R=4.1$ (41 kpc) about twice the radius of the disk.  
The choices of $L_{min}$ and $L_{max}$ limit the dynamic range of masses 
to about 600 with the least massive particles weighing in at 0.5\% 
the equivalent mass for a single-mass model and the most massive particle 
weighing in at $3\times$ the equivalent mass.  For comparison, 
the single-mass particle in the $N=10^8$ halo is 
$3\times 10^3$~\msolar, while in the multi-mass model, the particle 
masses range from 16 \msolar~for small $L$ to $10^4$ \msolar~for the 
most massive particles in outskirts of the halo.

\begin{figure}
\begin{center}
\epsscale{0.9}
\plotone{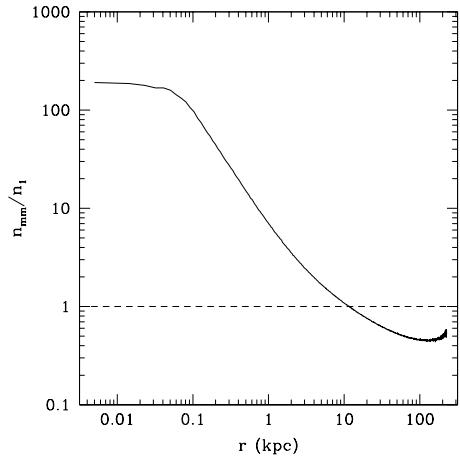}
\figcaption{
The ratio of the number density of the multi-mass 100M particle halo  to the
single-mass 100M halo.  The distribution function is sampled such that
particle mass is weighted by a smoothed step function of total angular momentum.
Particles with low angular momentum have small mass and those with high
angular momentum have low mass (see text).  The particle density is more
than 100 times higher within 0.1~kpc and at least 10 times higher within 1
kpc.  For $R>10$~kpc the number density drops gradually to about half
the single-mass case.  The effective numerical resolution at the
center of simulation is therefore $N_h\sim 10^{9-10}$.
\label{fig-nden}
}
\end{center}
\end{figure}

We plot the ratio of the particle number density in the multi-mass model to
the equal mass particle number density in Figure~\ref{fig-nden}.   
The number density is about $200\times$ greater within the central 
100~pc of the model and about $10\times$ at $R=1$~kpc.

\subsection{Simulations}

We simulate these models using a parallelized treecode \citet{dub96} for
200 time units (9.4~Gyr), permitting us to see the development of the bar
instability through various phases roughly over a Hubble time.  We soften
gravity with a Plummer model kernel and vary the softening length 
$\epsilon$ according to the particle numbers of the  
simulation roughly in proportion to $N^{-1/3}$.  The median force errors
are ~0.1\% for the chosen treecode parameters.
Simulation parameters are given in Table~1.

\begin{deluxetable}{llllll}
\tablewidth{0pt}
\tabletypesize{\footnotesize}
\tablecaption{Simulation parameters \label{tab-simulations}}
\tablehead{ \colhead{Model} & \colhead{$N_{h}$} &
  \colhead{$N_{d}$} & \colhead{$\epsilon$ (pc)} &
  \colhead{$\delta t$ (kyr)} & \colhead{$N_{steps}$} \\
}
\startdata
m10K & $10^4$ & $1.8\times 10^3$ & 200 & 470  & 20000 \\
m100K & $10^5$ & $1.8 \times 10^4$ & 100 & 470 & 20000 \\
m1M & $10^6$ & $1.8 \times 10^5$ & 50 & 470  & 20000\\
m10M & $10^7$ & $1.8 \times 10^6$ & 20 & 470  & 20000\\
m100M & $10^8$ & $1.8 \times 10^7$ & 10 & 470  & 20000\\
mm100M & $10^8$ & $1.8 \times 10^7$ & 10 & 235  & 40000\\
\enddata
\tablecomments{The model mm100M is the multi-mass model.}
\end{deluxetable}

A constant timestep is used for all of the simulations (see Table~1).
The circular orbital period in the mass model at the smallest softening radius
of $\epsilon=10$~pc is about 15~Myr and so is resolved by 30 timesteps. 
Plummer softening smooths gravity over a few softening lengths 
so the smallest resolved radius of these simulations is $\approx 3\epsilon$.  
We find that total binding energy is typically conserved to within 0.2\% 
and angular momentum is  conserved to within 1\% over the course of the 
runs.  Each simulation, with the exception of the multi-mass case, is 
repeated twice with a different random realization to explore  
statistical variance in the growth of the bar mode.

\begin{figure*}
\begin{center}
\plotone{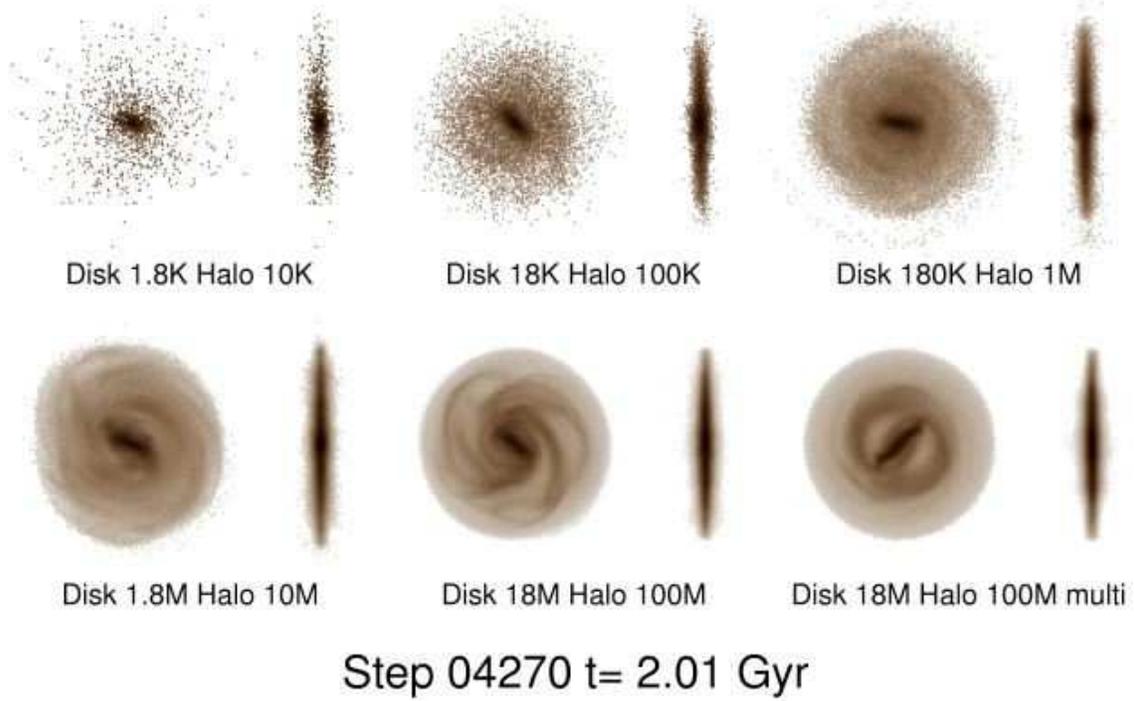}
\figcaption{
A comparison of the evolution of the bar instability in 6 simulations 
with increasing particle number $N$.  The formation of a bar is delayed for
simulations with larger $N$ since the Poisson seed noise has a 
lower amplitude and it takes longer for the instability to grow in the 
linear regime.  The lowest resolution simulations suffer from heating while
the general behavior converges at higher resolution for $N \ge 10^6$ (see
Video 1) 
\label{fig-animation1}
}
\end{center}
\end{figure*}

Figure~\ref{fig-animation1} shows an animation\footnote{Quicktime
animations are available at the website
www.cita.utoronto.ca/$\sim$dubinski/BarsInCuspyHalos/}
of the evolution of the 
disks in six models in face-on and edge-on views.
The lowest resolution model m10K demonstrates how insufficient particle 
numbers can lead to spurious results.  A bar develops immediately in the 
1.8K particle disk but  devolves into a compact rapidly tumbling object.  
In retrospect, early galaxy formation simulations that introduced 
the angular momentum problem \citep[e.g.,][]{nav00} only contained 
2K particles, so part of the problem may have arisen from exceedingly
noisy evolution of a bar mode.  The m100K model with an 18K particle disk 
still appears noisy, though the buckling instability is clearly visible.  
The disk is visibly thicker than the higher resolution models,  however, 
and the bar is not as pronounced.  Disk heating by bombardment of halo 
particles is a problem.
The time of onset of the bar instability is 
delayed as $N$ increases, reflecting the effect of Poisson noise.  
Since the bar instability grows exponentially from density fluctuations 
in the initial conditions, larger $N$ simulations will have smaller initial
amplitudes and, therefore, longer times to saturate.

\begin{figure*}
\begin{center}
\plotone{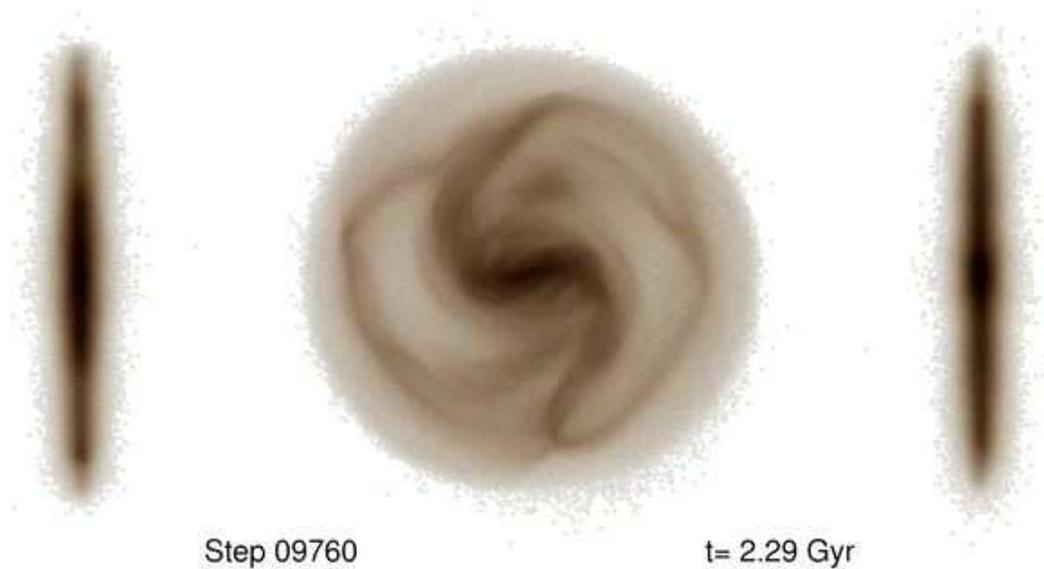}
\figcaption{
Evolution of the multi-mass model in the inertial frame showing the face-on
view and two perpendicular edge-on views.  The bar grows from the inside
out first evolving into a thin bar extending to the co-rotation radius and
then settling down into a less elongated ellipsoid.  The buckling instability 
vertically thickens the bar 
into a peanut-shaped bulge at later times.  The bar grows
in length as angular momentum is lost to the halo and new orbits are
captured with the co-rotation radius (see Video 2).
\label{fig-animation2}
}
\end{center}
\end{figure*}

\begin{figure*}
\begin{center}
\plotone{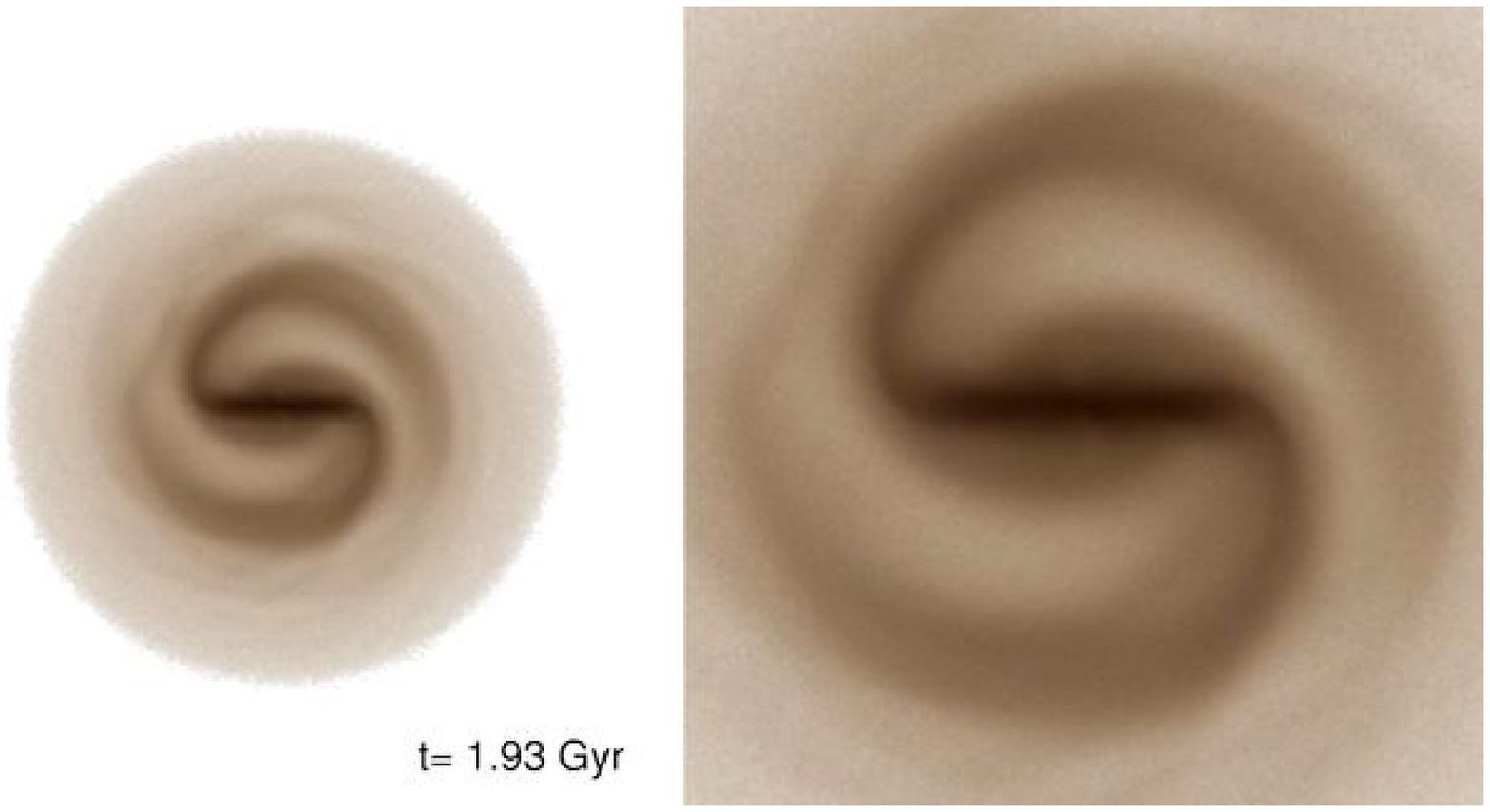}
\figcaption{
Evolution of the multi-mass model in the co-rotating frame showing a global
and close-up of the face-on orientation.  The co-rotation radius is clearly
visible at the distance where particles reverse the direction of
circulation around the bar.  The bar evolves to extend to the 
co-rotation radius and remains ``fast" (see Video 3).
\label{fig-animation3}
}
\end{center}
\end{figure*}

Figure~\ref{fig-animation2} displays the evolution of the multi-mass 
halo model with the 18M particle disk close-up and two perpendicular 
edge-on views simultaneously. Figure~\ref{fig-animation3}  refers to
the face-on view in a frame co-rotating with the bar to emphasize the 
growth of the bar mode.  This model starts very quietly and there is 
little visible structure until $t\approx 1$~Gyr when the bar begins 
to emerge.  The bar grows from the inside out, gradually increasing in 
length until it reaches a maximum length at nearly the CR radius 
around $t=2$~Gyr.  
At this time, the bar also excites a prominent, bi-symmetric spiral 
structure.  After saturation, the bar re-structures itself, becoming more  
centrally concentrated and weakening, as it settles into a quasi-steady 
state.  After settling, the pattern speed begins to
decline and the bar's length increases slowly, since the CR radius 
is increasing and the bar can capture additional orbits in the disk. 
The other notable event is the vertical buckling  instability that
occurs around $t=3.5$~Gyr creating a characteristic X-shaped structure
as various families of orbits establish themselves causing the bar to
thicken vertically.  By the end of the simulation, the inner bar transforms 
into a peanut-shaped bulge though it still is obviously elliptical in 
the face-on view.

Figure~\ref{fig-vrot-final} shows the rotation curve decomposition for the model at the
final time $t=9.4$ Gyr.   We rotate particles in the final barred disks to random 
angles $\phi$ to make the disk potential axisymmetic and then compute the disk rotation 
curve component through  $v_d^2=R\partial \Phi/\partial R$.  The halo rotation curve 
component is estimated from the spherically-averaged density profile of the dark matter.  The
collapse of the bar leads to a concentrated bulge-like component and the rotation curve 
flattens slightly creating a galaxy model that more closely resembles real systems. 
The halo profile at large radii does not change a lot but we will see
in further analysis discussed below that there is slight increase in the central
density.  In this barred galaxy model, both the stars and dark matter have comparable
contributions to the rotation curve in the inner regions.

In the next section, we quantify these various effects and look for
differences in resolution with the hope of finding numerical convergence in
physical behavior.

\section{Bar Growth and Pattern Speed Evolution}

The growth of the bar instability is measured by the bar strength 
with the $m=2$ Fourier amplitude of the surface density, $A_2$, given by:
\begin{equation}
A_2 = \frac{1}{M}\displaystyle\sum_{j=1}^N m_j \exp(2i\phi_j); R<R_c
\label{eq-a2}
\end{equation}
where the summation is performed over a list of particles with
masses $m_i$ at angle $\phi_i$ in the $x-y$ plane within some cut-off radius $R_c$.
The  normalized amplitude  $|A_2|$ versus time measures
the growth rate of the bar instability and the phase angle
$\phi = 0.5 \tan^{-1}[{\rm Im}(A_2)/{\rm Re}(A_2)]$
with time permits  
measurement of the pattern speed by numerical differentiation.

All models were simulated for 200 time units (9.4~Gyr) and every 
10 (or 20 in the multi-mass case) timesteps a face-on surface density 
image was generated from the particle distribution resulting in a sequence 
of 2000 images.  These  image sequences were analyzed to determine 
$A_2$ by summing over pixels rather than particles in equation \ref{eq-a2}.  
The amplitudes and phase angles were then tabulated as a function of time 
to determine the rate of growth of the bar and the pattern speed evolution.  
Pattern speed is  estimated by simply differencing angles in subsequent 
pairs of phase angles and dividing by the time interval.  In practice, 
we only use values every 50th snapshot to smooth out the noise in 
these parameters introduced by limited numbers of particles.  
We show below that higher resolution simulations produce smoother curves 
of bar growth and pattern speed evolution.

\subsection{Numerical Accuracy}

We first discuss the behavior of the bar instability as a function of integration timestep.
\citet{kly08} have claimed that very small timesteps are necessary to resolve
the dynamics of bars because of the possible development of cuspy density profiles in
the forming bulge-bar system.  Our simulations use a single timestep chosen to resolve 
the smallest dynamical timescale in the model.
Multiple timestepping schemes often use the criterion, 
$\delta t = (2.8 \epsilon/|{\bf g|})^{1/2} \eta$ where $\epsilon$ is the Plummer softening or
equivalent, $|{\bf g}|$ is the acceleration and $\eta$ is a free parameter usually chosen with a
recommended value of $\eta=0.2$ \cite[e.g.,][]{spr01}.  
For density laws following $\rho \sim r^{-1}$ the central
acceleration is constant.  The highest value of the acceleration in our galaxy 
model occurs in the center, and using it with $\eta=0.2$ we arrive at 
$\delta t=0.01$ (470 kyr) for $\epsilon=50$ pc and $\delta t=0.004$ (190 kyr) for 
$\epsilon=10$ pc.  Plummer softening of course reduces the maximum value near 
the center and it formally falls to zero at $r=0$.
We see below that there is only a modest increase in the central 
density evolution, so  the maximum value of $|{\bf g}|$ does not change 
by much over the course of the run.

The smallest orbital period is $\sim 15$ Myr for an orbit with $R\sim \epsilon$ and our
chosen timestep is $\delta t = 470$ kyr, so these orbits are resolved with approximately 30
timesteps.  For our highest resolution simulation, we use $\delta t = 235$ kyr 
to account for the smaller softening radius.  The fraction of particles with orbital 
periods less than 20 Myr is approximately 0.1\% based on a analysis of the radial 
frequency of 100K testparticle orbits sampled from the halo integrated
within the rigid potential of our mass model.  If the timestep is too large, orbits near the
center will be unstable and create an artificial constant density core.
Another possible problem occurs for highly radial orbits with longer periods that pass close to the
central cusp.  Our orbital analysis showed that approximately 0.13\% of orbits change binding
energy by more than 1\% over a 9.4 Gyr integration with $\delta t=470$ kyr.  All of these orbits
had small pericentric radii $\sim 100$~pc.   We therefore
expect a small fraction of highly radial orbits to diffuse artificially through energy space.  
We demonstrate here that the single timesteps
of $\delta t = 235$ and 470 kyr are sufficiently small to resolve the dynamics
for our choices of the Plummer softening radius.

To test for numerical convergence, we have re-run the model m1M using {\em single} timesteps
over the range of $\delta t=15-940$ kyr for a time of 4.7 Gyr.  The model with $\delta
t=15$ kyr required 320K single steps.  This galaxy model has
$N_d=180$K and $N_h=1$M and a Plummer softening length of $\epsilon=50$ pc. 
We examine different metrics of the system evolution including bar growth, pattern speed 
evolution and the final density profile of both the stars and DM all as a function of
timestep.

\subsubsection{Acceleration Errors}

\begin{figure}
\begin{center}
\plotone{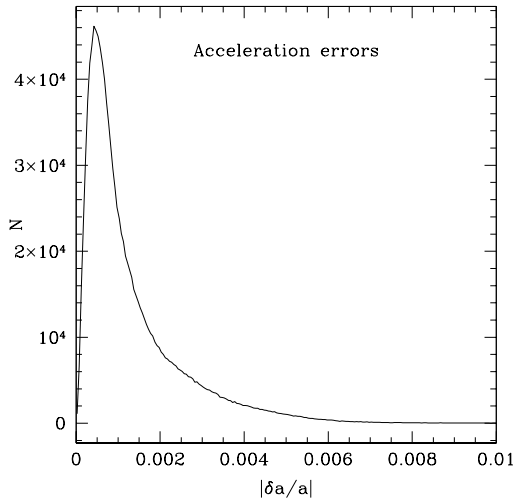}
\figcaption{
Relative acceleration errors for the parallel treecode for runs with
$N=1.18$M particles.
Errors are estimated by comparing acccelerations computed with our
preferred treecode opening angle parameter $\theta=0.9$ with quadrupole
order corrections to the exact values determined from a direct
calculation.  The mean relative error is 0.13\% with a median value of
0.085\%.  Note the opening angle criterion for the parallel treecode is
more conservative than standard definitions and so a larger value of
$\theta$ still results in relatively small acceleration errors (Dubinski
1996).
\label{fig-acc-err}
}
\end{center}
\end{figure}

We first comment on the accuracy of the accelerations determined using the parallel
treecode (Dubinski 1996).  Normally force accuracy is not discussed despite a large
variety of algorithms used to compute gravitational forces.  We present our errors here
so that other researchers may compare to their own standards of numerical accuracy.
Figure~\ref{fig-acc-err} shows the distribution of relative acceleration
errors for our preferred treecode parameters.  We use an opening angle
tolerance $\theta=0.9$ with 
quadrupole corrections using a more conservative cell opening criterion than normally described
that gives more accurate acceleration values for a given $\theta$ than
standard definitions (Dubinski 1996).  Errors are determined by comparing
accelerations from the treecode method to a direct force calculation.  The median and
mean relative acceleration errors are 0.085\% and 0.13\% respectively with 99.7\%
($3\sigma$ limit) of acceleration errors less than 0.7\%.

\subsubsection{Energy Conservation}

\begin{figure}
\begin{center}
\plotone{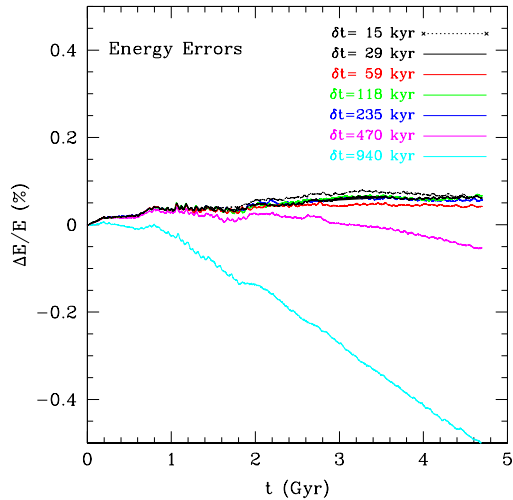}
\figcaption{
Total energy errors for runs with different time-steps $\delta t$.  The simulation with
$\delta t = 940$ kyr shows a systematic drift due to inadequate numbers of 
timesteps to follow orbits within the core.   There is a lesser drift for the timestep
$\delta t=470$ kyr but the error has only grown to 0.1\% by the end of the run.  All
timesteps with $\delta t < 470$ kyr show very little drift.
\label{fig-energy}
}
\end{center}
\end{figure}

The evolution of the error in total binding energy of an N-body system is a useful indicator 
of the fidelity of the results and can reveal potential problems with the integration
scheme or choice of timestep.  Figure~\ref{fig-energy} shows the change in total binding
energy as function of timestep.  The largest timestep of $\delta t=940$ kyr shows a
strong systematic drift in energy reflecting the inadequate timestep 
resolution for a significant fraction of orbits.  There is a smaller drift in the energy with
a relatively small error of 0.1\% over 4.7 Gyr with our main 
choice of $\delta t=470$ kyr and clear convergence with no systematic effects 
with $\delta t \le 235$ kyr. We, therefore, conclude that $\delta t = 470$ kyr is 
adequate for our models.  We show below that there are no substantial differences 
between various metrics of the properties of the bar and halo when using timesteps 
with $\delta \le 470$ kyr.

\subsubsection{Bar Evolution versus Timestep}

\begin{figure}
\begin{center}
\plotone{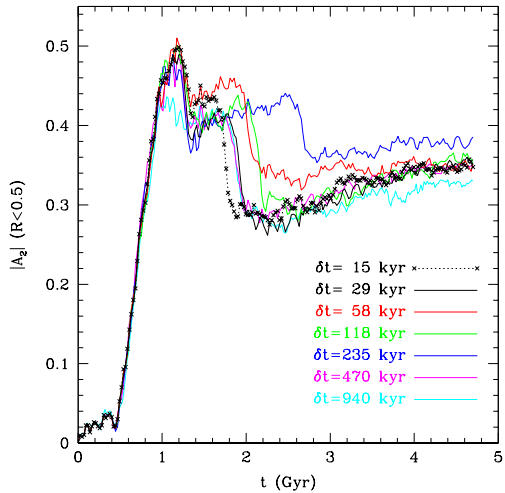}
\figcaption{
Evolution of the Fourier component, $A_2$ for stars with $R<5$~kpc for a
single model with $N_d=180K$ and $N_h=1M$ particles with global timesteps 
spanning the range of $\delta t=15$ kyr to
940 kyr.  The linear growth phase of the bar is almost identical until the bar
instability saturates at $t \sim 1$ Gyr.  The subsequent nonlinear
evolution shows a wide range of behavior for different timesteps with 
no monotonic trend.  The nonlinear phase of the bar instability involves 
chaotic orbits and so the slight variations introduced by the round-off error of 
different discrete timesteps lead to divergent evolutionary behavior.
The main manifestation of this chaos are different times for the onset of
the buckling instability ranging from 1.8-2.5 Gyr having no dependence on the
chosen timestep.   Nevertheless, the behavior is qualitatively similar
after the buckling instability with a steady rise of $|A_2|$ at late times
as the bar lengthens.
\label{fig-a2-dt}
}
\end{center}
\end{figure}

We measured both bar growth and pattern speed evolution as a function of timestep in the
m1M model with $N_d=180$K and $N_h=1$M.  
Figure~\ref{fig-a2-dt} presents the evolution of the bar growth parameter $|A_2|$ measured 
within $R<5$~kpc versus timestep.  During the linear growth phase of the bar 
instability, all simulations track one another very closely.  However, after the bar
instability saturates around $t\sim 1$ Gyr the behavior is quite variable and erratic
for different choices of the timestep.  The time of bar buckling shown by the sudden
secondary drop in $|A_2|$ changes with different timesteps and lies in the range
$t=1.8-2.5$ Gyr.  There is no monotonic trend with timestep.  The range of variability
is the same as our study of independent random realizations in \S~\ref{sect-fixed}.
The root cause of this
behavior is probably the dynamical chaos inherent to this late evolution of the bar
instability.  
The detailed $N$-body solutions for individual particles diverge exponentially for
different choices of integration step in the nonlinear regime of dynamical evolution.  
Despite this divergence on the individual particle level, the global properties of the
resulting bar are similar as we shall see.

\begin{figure}
\begin{center}
\plotone{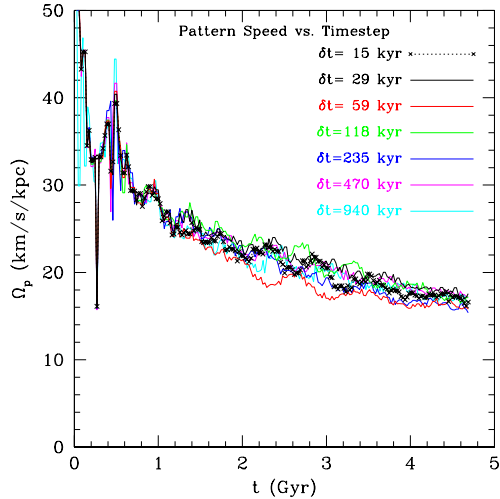}
\figcaption{
Evolution of the pattern speed $\Omega_b$ for models 
with $N_d=180$K and  $N_h=1$M and timesteps 
spanning the range of $\delta t=15$ kyr to
940 kyr.  During the linear growth phase of the bar until $t\approx 1$ Gyr, 
the pattern speed evolution is almost the same.  Once the bar becomes
nonlinear, there is a small scatter in the detailed behavior of the 
pattern speed with a variation of $2 - 4~\kmskpc$ at any given time.
By the end of the runs, the results converge with the mean and variance 
of the pattern speed $\Omega_b = 16.6 \pm 0.7~\kmskpc$ at $t=4.7$ Gyr.
There is no strong dependence of the pattern speed evolution on the choice
of timestep.
\label{fig-omega-dt}
}
\end{center}
\end{figure}

An analysis of the pattern speed evolution shows consistent results for all timesteps
(Fig.~\ref{fig-omega-dt}).  The agreement in the linear regime evolution until the 
bar instability saturates at $t\sim 1$ Gyr is very close, after which the detailed
evolution show differences.  There is a $2-4~\kmskpc$ scatter in the pattern speed at
any given time but the general declining trend is the same over the course of the run.  
The observed scatter is consistent with the same scatter seen in different random
realizations (Fig.~\ref{fig-1M_seq-a2_0.5}).  The mean and variance of the pattern speed
at $t=4.7$ Gyr for all timestep runs is $\Omega_b = 16.6 \pm 0.7~\kmskpc$.  We conclude
that our choice of timestep leads to a consistent evolution of the bar pattern speed.

\subsubsection{Stellar and Halo Central Density versus Timestep}

As a final metric of the accuracy of the simulations versus timestep, 
we measured the spherically
averaged density profile of the stars and DM at the last snapshot at $t=4.7$
Gyr. At this time, the bar has buckled and has formed a concentrated bulge-like object
within the halo that has become more dense itself in response to this new
bulge (see below).  Figure~\ref{fig-dden-dt} shows the stellar density profile within $r<1$~kpc 
of the center for runs with different timesteps.  The density profiles for the
bar/bulge are consistent within the error bars for $\delta t \le 470$ kyr.  There is
some random scatter in the inner radial bins since there are only a few hundred
particles at these small radii.  The model with $\delta t=940$ kyr forms a core with
constant density within $r < 200$ pc though the density is only 0.3 dex (about
$2\times$) smaller than the density in the first radial bin of the smaller 
timestep runs.  The runs with
$\delta t \le 470$ kyr agree within $\pm 0.1$ dex for $r<200$ pc and much of that error
is due to small particle numbers with $\sim 10^2$ particles per bin.

We also measured the spherically-averaged density profile of the DM at $t=4.7$
Gyr (Fig~\ref{fig-hden-dt}).  We have approximately $6\times$ as many particles per bin and
so the random errors are smaller.  The DM density profiles are consistent for
$\delta t \le 470$ kyr suggesting we have adequate time resolution for the halo
density evolution.  Again, we see the development of an artificial
constant density core in the simulation with $\delta t = 940$ kyr.  This timestep is
clearly too large and does not adequately follow short period orbits in the core.  
However, simulations with timesteps smaller than $\delta t \le 470$ kyr adequately follow the
dynamics of the evolution of the DM halo.

\begin{figure}
\begin{center}
\plotone{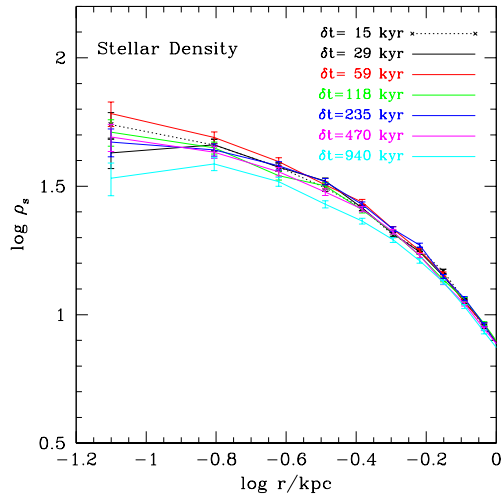}
\figcaption{
The spherically averaged density profile of the stellar component that
includes the buckled bar and disk at $t=4.7$ Gyr for runs with different 
time-steps.  The error bars are $1-\sigma$ estimates of 
the $\sqrt{N}$ Poisson error in the
density due to discrete sampling e.g., the inner most bins contain
$\sim 100$ particles so the $1-\sigma$ error in density is about 10\%.
For time-steps with $\delta t \le 470$ kyr,  the density profiles are
consistent within the random errors.  The run with $\delta t = 940$ kyr
shows the formation of an artificial core due to an insufficient number of 
time steps to follow orbits within $r \sim 100$ pc.
\label{fig-dden-dt}
}
\end{center}
\end{figure}

\begin{figure}
\begin{center}
\plotone{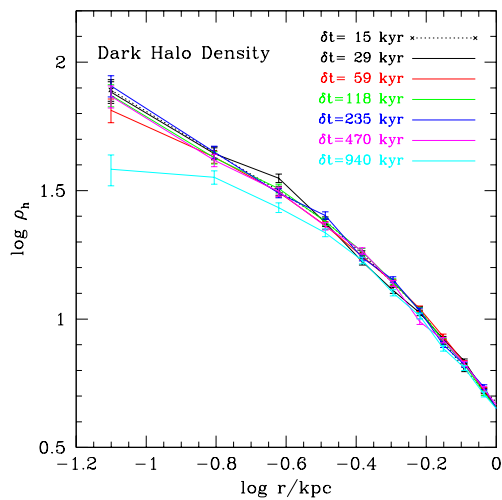}
\figcaption{
The spherically averaged density profile of the dark matter halo
at $t=4.7$ Gyr for runs with different 
time steps.  The error bars are $1-\sigma$ estimates of 
the $\sqrt{N}$ Poisson error in the
density due to discrete sampling.
For time-steps with $\delta t \le 470$ kyr,  the density profiles are
consistent within the random errors. The run with $\delta t = 940$ kyr
shows the formation of an artificial core due to an insufficient number of 
time steps to follow orbits within $r \sim 100$ pc.
\label{fig-hden-dt}
}
\end{center}
\end{figure}

In summary, we have presented the force accuracy and total energy evolution of our
simulations with different timesteps.  We have also shown that our results converge
experimentally for $\delta t \le 470$ kyr
according to different metrics of the bar evolution including pattern speed evolution
and stellar and dark matter density central density profile.  We note that there is a
random behavior for the time of onset of the buckling instability for different choices
of timestep which probably results from the chaotic nature of this dynamical system.
(This was shown explicitly by Martinez-Valpuesta \& Shlosman 2004.)
If the timestep is too large, the main effect is to create an artificial constant 
density core.  Particles with short orbital periods are numerically unstable and are
scattered out of the center creating the core.  In the subsequent analysis, we show 
that the central density continues to increase at smaller radii with higher mass resolution.  If our
timestep was too large, one might expect instead to see the onset of a artificial constant 
density core of a fixed radius set by the timestep and independent of the mass resolution.
We do not observe this behavior.  We also do not see a sudden change in behavior of the pattern
speed evolution at a critical timestep as seen by \citet{kly08}.
We, therefore, conclude that we have adequate time resolution to follow the dynamical 
evolution of this system all the way down to the radius where Plummer softening dominates.

\subsection{Models at fixed resolution \label{sect-fixed}}

Before presenting results on the bar and pattern speed  evolution versus
mass resolution, it is instructive to understand the variance expected 
for runs at a fixed resolution.  The seed of both spiral and bar 
instabilities in $N$-body simulations is the Poisson noise in the 
discrete particle distribution of the disk and  halo.  We, therefore, 
expect some variation in the detailed behavior of the growth of the bar 
instability in different random realizations and we quantify it here.

\begin{figure}
\begin{center}
\plotone{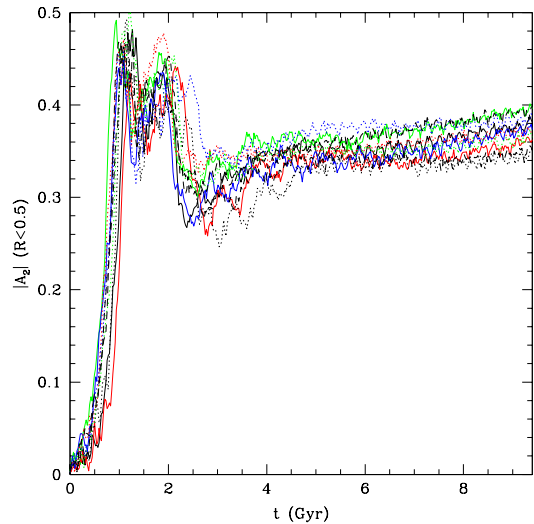}
\figcaption{
Evolution of the Fourier component, $A_2$ for stars with $R<0.5$ for  10 models
with $N_d=180K$ and $N_h=1M$ generated with different initial random
seeds.  There is a large variation in evolution of $A_2$ during the 
formation of the bar over the time interval $t=20-70$ reflecting
detailed differences in the Poisson noise in different random realizations.
The plot reveals the approximate scatter in evolutionary behavior expected 
for different runs.
\label{fig-1M_seq-a2_0.5}
}
\end{center}
\end{figure}

We build ten galaxy models with 1M halos particles and 180K disk
particles independently from different Monte-Carlo samplings of the
galaxy model DF by using a different initial seed for the random number
generator.  We measure $|A_2|$ within a radius $R<0.5$ units (5~kpc)  
which is within the eventual co-rotation radius of the bar.  Figure
\ref{fig-1M_seq-a2_0.5} shows the evolution of the bar strength
for the 10 runs.  The detailed behavior varies significantly for
the different runs with the minimum and maximum values of $|A_2|$
that varying by $\pm 0.05$ during the bar growth phase between $t=1-3.3$ Gyr and
final values ranging from 0.35-0.40 at $t=9.4$ Gyr.  While the runs differ
in detail there is still a generic behavior with the bar growth  saturating
around at $|A_2| \approx 0.5$ and then going through an oscillation  before
settling down to a value near $|A_2|\approx 0.3$ around $t=3.3$ Gyr.
The bar then grows slowly increasing in length as the pattern speed
declines.

\begin{figure}
\begin{center}
\plotone{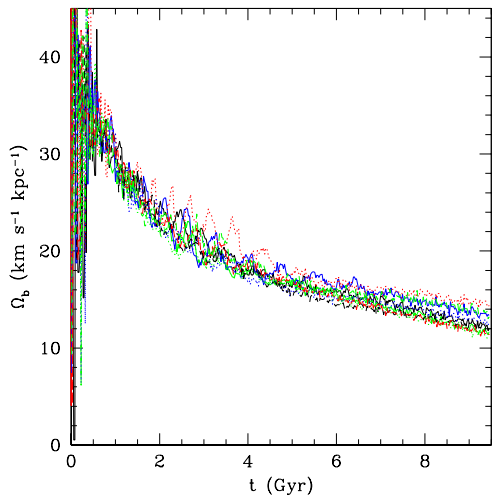}
\figcaption{
Evolution of the pattern speed, $\Omega_b$, for 10 models with $N_d=180K$
and $N_h=1M$ generated with different initial random seeds.
The pattern speed is measured by creating a time series of the phase angle
of the $A_2$ component for stars with $R<0.5$.   The pattern speed decays
after the bar forms as angular momentum is transferred to the dark 
halo through dynamical friction.  While the decay rate is similar, 
again there is scatter due to statistical variation of the Poisson noise 
in the initial conditions.
\label{fig-1M_seq-omega_0.5}
}
\end{center}
\end{figure}

Fig. \ref{fig-1M_seq-omega_0.5} shows the pattern speed $\Omega_b$
evolution for the same 10 runs at fixed resolution.  The behavior is
similar with the bar starting out with 
$\Omega_b \approx 35 \kmskpc$
declining to a value between $12-14 \kmskpc$.  
The pattern speed  evolution
is consistent at the 10\% level despite the different histories of the bar
growth as quantified by $|A_2|$.

We, therefore, expect the minimum and maximum values of $|A_2|$ to vary by
about 0.05 units between models and pattern speeds to vary conservatively 
by $\pm 2 \kmskpc$ for stochastic reasons alone.

\pagebreak

\subsection{Models with increasing mass resolution}

After quantifying the effects of temporal resolution and
stochasticity in models at fixed resolution, we go on to
examine models of increasing mass resolution with halos containing from $10^5$
to $10^8$ particles and the multi-mass model with an effective  
resolution of $10^{10}$ particles.  Our goal here is to measure 
carefully bar growth and pattern speed and to single out any differences that 
are inconsistent with the expected statistical variance.  We have simulated 2
models at each resolution in Table 1 with the exception of the multi-mass
case where we did only one model.

\begin{figure}
\begin{center}
\plotone{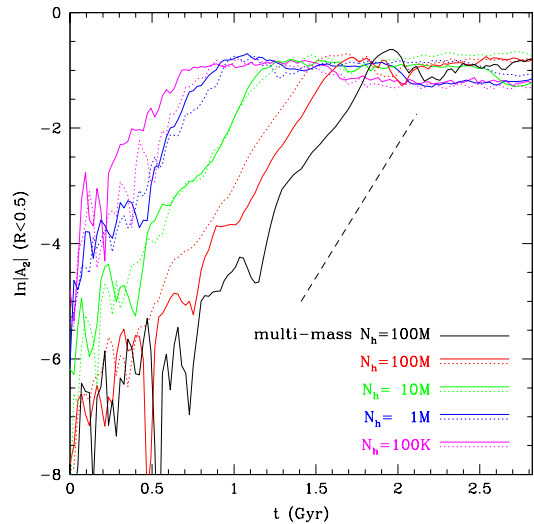}
\figcaption{
Initial growth of the bar strength $|A_2|$ for stars with  $R<0.5$
for two model sequences using 
$N_d=18K,180K,1.8M,18M$ with $N_h=100K,1M,10M,100M$ respectively.
The $\ln|A_2|$ grows approximately linearly with time independent of the 
choice of $N_d$ and $N_h$ showing the exponential growth of the bar mode.  
The dashed line shows an exponential timescale that is approximately 
$\tau=370$~Myr. Since the bar grows from the Poisson noise within the 
disk then we expect the noise amplitude to be proportional to $N^{-1/2}$.  
Based on exponential growth of the bar mode, we expect the time to 
saturation of $|A_2|$ to be delayed by roughly 
$\delta t\approx \tau \ln(N_1/N_0)^{1/2}$, e.g., 
a factor of 10 change in particle numbers leads to a delay 
$\delta t  \approx 9$.  The difference in saturation times of $|A_2|$ 
between the various simulations are roughly consistent with this 
estimate though  there is some variation.
\label{fig-lseq-all-a2_0.5}
}
\end{center}
\end{figure}

Figure \ref{fig-lseq-all-a2_0.5} shows the bar growth for all
resolutions plotted as $\ln|A_2|$ versus time to emphasize the growth of the
instability through the linear regime.  Spiral and bar instabilities grow
from seed density fluctuations in Poisson noise through the swing
amplification mechanism \citep{too81}.  In the linear regime, the
fluctuations grow exponentially and so $\ln|A_2|$ is roughly linear in time.
The dashed line to the right is parallel to the model growth rates and
corresponds to exponential growth with a timescale of $\tau=370$~Myr.

Once the perturbation goes non-linear, $|A_2|$ reaches a maximum value and
then oscillates until reaching a steady state as the bar settles into a 
quasi-equilibrium.  All models show a gradual linear rise of
$\ln|A_2|$ after reaching equilibrium but are noticeably offset in
the saturation time when going to higher resolution.  Since the seed 
perturbations arise from Poisson noise, the amplitude of perturbations 
$\delta$ varies as $N^{-1/2}$, so the ratio of amplitudes in two different 
simulations is $\delta_1/\delta_0 = (N_1/N_0)^{-1/2}$.
In the linear regime, $\delta \sim \exp(t/\tau)$, so the time  
delay between growing perturbations to reach the same amplitude is 
$\delta t  \approx \tau \ln(N_1/N_0)^{1/2}$.  Simulations with a factor of 
10 more particles will, therefore, be delayed in saturating by a time 
interval given by $\delta t \sim \tau \ln 10^{1/2}\approx 1.1\tau$.  
With $\tau \approx 370$~Myr, we expect a time delay of approximately 400 Myr
between simulations differing by a factor of 10 in numbers of particles.  
If we select the time when $|A_2|$ reaches a maximum 
as a reference time when the bar saturates and the linear regime ends, we 
can estimate the time delay between simulations directly.  Using the 
1M particle run as a zero point,  we find delay times of 
$\delta t=280$~Myr for 10M particle models and $\delta  t=600-700$~Myr for the
100M models and $\delta t=950$~Myr for the multi-mass 100M model.  The noise
characteristics of the multi-mass model are more complicated than the
simple ideas discussed here and vary across the model but the onset 
of the bar instability is nonetheless delayed further because of 
quieter initial conditions. These values are slightly smaller than expected 
but are in reasonable agreement with the estimated delays from  
considerations of the growth of Poisson fluctuations.  This analysis 
emphasizes that {\em the spiral and bar instabilities that arise in 
N-body simulations of  disks are wholly dependent on the initial Poisson noise.
In the future,  with simulations using more than 10M
disk particles it makes sense to control the properties of the noise
both in amplitude and power spectrum as done in cosmological  
simulations.}

\begin{figure}
\begin{center}
\plotone{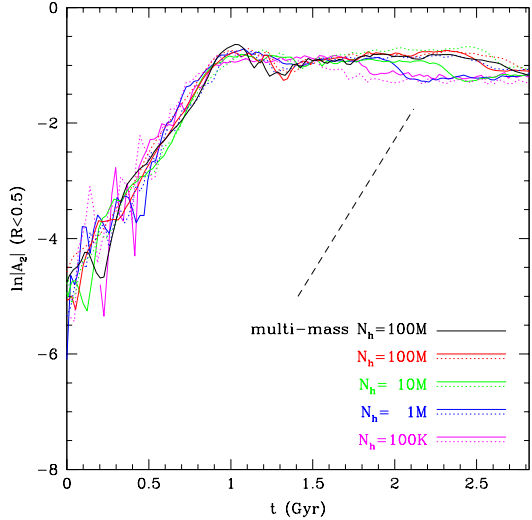}
\figcaption{
Same as Fig. \ref{fig-lseq-all-a2_0.5}
except the curves have been shifted
in time so that the linear regime growth phases overlap with 
the m1M model according to the measured time delays
\label{fig-lseq-all-a2_0.5-offset}
}
\end{center}
\end{figure}

\begin{figure}
\begin{center}
\plotone{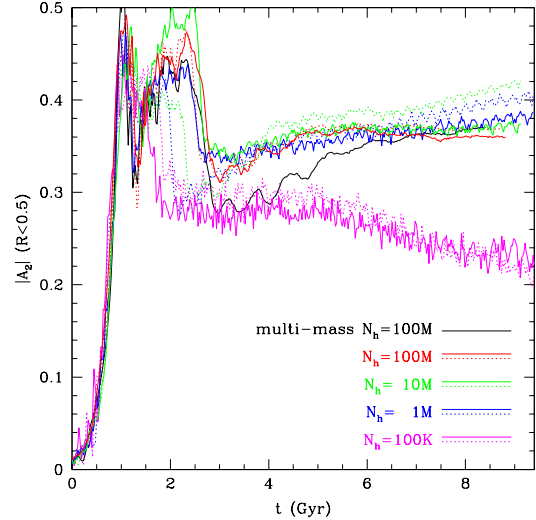}
\figcaption{
Evolution of the bar strength $|A_2|$ for stars with $R<0.5$ for two 
model sequences using 
$N_d=18K,180K,1.8M,18M$ with $N_h=100K,1M,10M,100M$ respectively and the
multi-mass model plotted versus linear time.
The curves have been synchronized to the time of maximum bar extent.
This plot emphasizes the variance in behavior after the bar instability
goes nonlinear. 
\label{fig-seq-all-a2_0.5-offset}
}
\end{center}
\end{figure}

Figure \ref{fig-lseq-all-a2_0.5-offset} and \ref{fig-seq-all-a2_0.5-offset}
show the evolution of the bar strength when we account for the time delays
and allow a comparison of the early and late time evolution.  When
synchronized this way, the linear growth phase is readily apparent in
the evolution of $\ln|A_2|$ in Fig.~\ref{fig-lseq-all-a2_0.5-offset}.  
In Fig.~\ref{fig-seq-all-a2_0.5-offset}, the plot of the evolution of $|A_2|$
reveals the details of the non-linear evolution of the bar.  The bar
strength saturates at a maximum value followed by an oscillation
through a minimum and then a slow rise to the end of the simulation.   
For the most part, the range of behavior between different resolutions is
consistent with our expectations of variance from our study of 10
simulations at fixed resolution.  However, the highest-resolution
multi-mass model dips to a lowest minimum value of $|A_2|$ and takes more time
to grow in the later phase.  This difference does lie within the range of
stochastic behavior but still appears slightly anomalous.  We will
find below that the rate of angular momentum transfer between
the bar and the halo is slightly slower for the multi-mass run.
The multi-mass run seems to transport about 10\% less angular momentum
from the bar to the halo than the other runs and this could account  
for the different behavior.

\begin{figure}
\begin{center}
\plotone{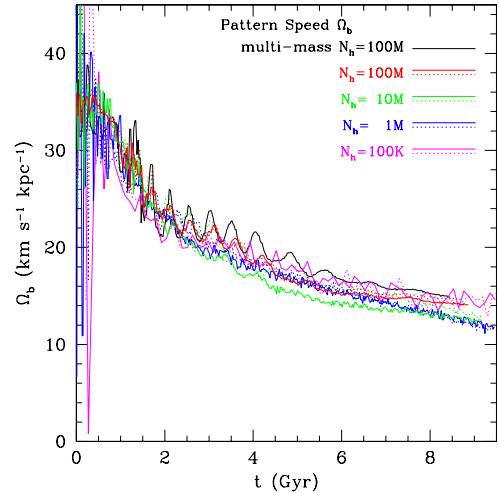}
\figcaption{
Evolution of the pattern speed $\Omega_b$ for two model sequences using
$N_d=18K,180K,1.8M,18M$ with $N_h=100K,1M,10M,100M$ respectively  
($R<0.5$).  Also,
shown is one model with $N_d=18M$ and $N_h=100M$ with a multi-mass halo
that increases the particle number density near the center of the disk.
The curves have been shifted in time so that the bar growth evolution 
is coincident with the m1M model.  The decline in pattern speed
at different resolutions is similar though there the multi-mass model
does not decay as quickly and has a slightly larger pattern speed at the
last simulated point.  The 100M particle simulations also show a modulation
of the pattern speed that indicates more subtle dynamical effects revealed
by higher resolution.
\label{fig-seq-all-omega-offset}
}
\end{center}
\end{figure}

Finally, we compare the pattern speed evolution of simulations at different
resolutions.  Figure \ref{fig-seq-all-omega-offset} shows the pattern
speed versus time for all simulations where again for a proper comparison
we have synchronized the various runs to the time of the first peak in 
$|A_2|$ as before.   The decline of the pattern speed is similar for 
all resolutions with bars initially forming with 
$\Omega_b \approx 35 \kmskpc$ and ending
with a value around $\Omega_b \approx 12-14 \kmskpc$.  
The range of
curves is again consistent with the scatter seen in the fixed resolution study.
The highest resolution runs with 100M halo particles in both the single
mass and multi-mass case show an apparent oscillation in $\Omega_b$
during the decline. The frequency of this oscillation is approximately half
of the pattern speed $\Omega_b$ itself.  The source of the oscillation is 
not obvious.  We initially speculated that interference from spiral patterns 
beyond the end of the bar rotating at a different pattern speed may have 
altered the measurement of $A_2$ within $R<0.5$.  However, when the pattern 
speed is derived from $A_2$ measured within $R<0.25$ out of influence 
of spirals the oscillations persist at the same frequency.  These oscillations 
may result from uneven bar growth (i.e., variations in length) 
by trapping of the disk orbits by the bar or from
a nonlinear mode coupling \citep{mar06,mar06b}.

In summary, the bar develops from Poisson noise in the disks in a similar
way for simulations with $N_h>10^6$. The time delay in the growth to the
nonlinear phase for larger $N_h$ are the result of smaller amplitude
Poisson fluctuations that seed the bar at higher resolution.  The variation
in the behavior of the different runs is consistent with the variance
introduced from different random realizations of the models.
The bar pattern speed decays at a similar rate over the course of
the run for resolutions again with $N>10^6$ though the higher resolution
runs decay to a final value that is approximately 10\% larger.  There  
is no dramatic change in dynamical evolution of gross physical properties  
of the bar as we approach $N_h=10^8$ suggesting the models are converging to  
the correct physical behavior.

\subsection{A Fast Bar}

Orbital dynamics permits a bar of length $a_b$ to extend as far  
as the CR radius $D_L$ \citep{con80}. 
But the developing chaos  
between the Ultra-Harmonic resonance (UHR) and the CR limits the bar  
length to within the UHR, especially in stronger bars.
The dimensionless ratio $\mathcal{R}=D_L/a_b$ is an indicator of a
bar's dynamical state and galaxies with observed or inferred pattern speeds
have $\mathcal{R} = 1.2\pm 0.2$ \citep[e.g.,][]{ath92,deb98}.
Bars emerging from the disk instability are usually born with
$\mathcal{R}\approx 1$ and this value gradually increases as the bar
settles into equilibrium and loses angular momentum to the halo through
dynamical friction. During buckling the bar shortens dramatically
for some period of time \citep{mar04,mar06} and afterwards
gradually lengthens.  However,  the CR radius also increases in  
response to the change in potential of both the outer disk and DM 
halo as they absorb angular momentum from the bar and respond 
to the changing mass profile of the disk. \citet{deb00}
have shown that in many models with dense halos, bars are slowed down
considerably and end up with values of $\mathcal{R}>2$.  Bars are,  
designated as ``fast" if $1<\mathcal{R} < 1.4$ or ``slow" for 
$\mathcal{R} > 1.4$ with all barred galaxies with determined or inferred 
pattern  speeds being ``fast" by this definition.

\begin{figure}
\begin{center}
\plotone{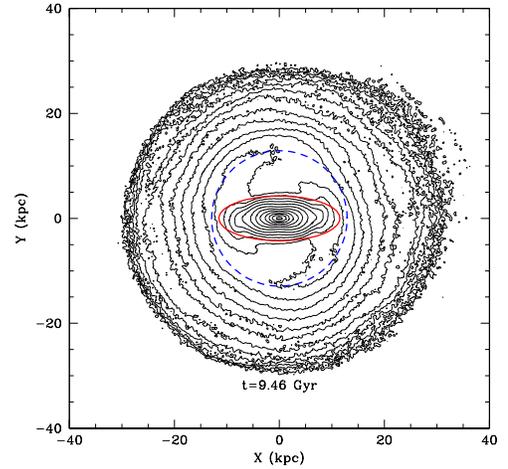}
\figcaption{
Surface brightness contours of the multi-mass model at the last snapshot
at $t=9.4$ Gyr overlayed with the best fit ellipse to the central bar 
and co-rotation
radius.  Even at this late time in the evolution, 
the bar extends to the co-rotation radius.
\label{fig-contour}
}
\end{center}
\end{figure}

\begin{figure}
\begin{center}
\plotone{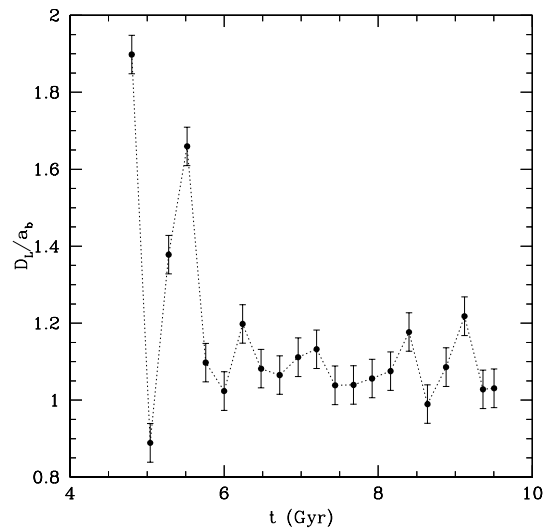}
\figcaption{
Evolution of the CR radius to bar length ratio $\mathcal{R}$ for the
multi-mass model over the last half of the simulation.  The value of
$\mathcal{R}$ hovers around 1.1 indicating a fast bar.
\label{fig-barratio}
}
\end{center}
\end{figure}

The bar that forms in the model described here is ``fast" with $ \mathcal{R}
\la 1.4$ after reaching a quasi-equilibrium after buckling.  
The result is in  agreement
with \citet{mar06} who used a similar galactic model and  determined the
bar size by means of the last stable orbit supporting it.  At  various times 
after the bar forms, we have determined the curve of the circular orbital 
frequency $\Omega(R)$ by computing the average of the radial acceleration 
$d \Phi/dR$ on points on rings of different radii in the midplane of the disk.
In this way, we average out the asymmetry in the potential introduced by
the bar.  The CR radius is then found by reading off the radius
corresponding to $\Omega(D_L)=\Omega_b$ at the given time.  To  
determine the bar length $a_b$ we fit elliptical contours to the surface 
density  profile and look for a sudden transition in the value of the 
axis ratio $q=b/a$ and the position angle of the isodensity contours.  
In most cases, the transition is sudden, jumping from $q=0.4$ to $q=0.9$ 
over a radial interval of 1 kpc.  We therefore can determine $a_b$ with an 
accuracy of $\pm  0.5$ kpc.  Figure~\ref{fig-contour} shows the isodensity 
contours overlayed with corotation radius and elliptical contour for 
the chosen bar  radius at $t=9.4$~Gyr the final time in the 
simulation for the highest resolution model mm100M.  Even at this time, the
bar nearly extends to the CR radius and $\mathcal{R} = 1.2 \pm  
0.05$.  Figure \ref{fig-barratio} shows the evolution of $\mathcal{R}$ from
$t=4.7-9.4$~Gyr starting with the time when it just has  settled
into equilibrium until the end of the run.  During this time, the bar  ratio
$\mathcal{R}$ maintains a value less than 1.4 and so would be classified as
a ``fast" bar and remain consistent with the observed barred galaxies.
Despite a 3-fold drop in the pattern speed the bar length and the  
galaxy potential readjusts to keep $\mathcal{R}$ near 1.
\citep{deb00} found that models with $V^2_{disk}/V^2_{halo}=1$ 
do end up with fast bars and indeed our model is consistent with that value 
(Fig.~\ref{fig-vrot1}).  

Finally, we estimate the bar mass and shape for comparison to current studies on
bar-halo interactions.   The best fit ellipsoid to the bar with  $a_b=12.5$
kpc has axis ratio $a_1:a_2:a_3 = 3.6:1.4:1$ The mass in the disk within
the elliptical contour of $q=0.4$ with bar length $a_b=12.5$ kpc is
$M_b=3.7\times 10^{10}$~\msolar~compared with a total disk mass of
$M_d=5.5\times 10^{10}$~\msolar.  The bar therefore represents 2/3 of the
total disk mass.  The mass of the DM halo within  the sphere of radius
$a_b=12.5$~kpc is $M_h(r<a_b) = 6.1\times 10^{10}$~\msolar~compared  
with a total halo mass $M_{h,tot} = 3.0\times 10^{11}$~\msolar.  
The ratio  of bar to total halo mass enclosed is $M_b/M_h(r<a_b)\sim 0.6$.
We will see below that this perturbation has a small effect on the 
density profile of the DM halo and is not sufficient to create 
a flat density core as seen in recent work with a rigid bar evolving in 
a spherical $N$-body halo with $M_b/M_h=0.5-1.0$ and a thinner bar with 
$a_1:a_2:a_3=10:2:1$ \citep{wei07b}.

\section{Halo Density Profile Evolution}

Our next task is to examine the evolution of the dark halo density profile.
\citet{wei02} originally demonstrated that a thin {\em rigid}  
bar rotating within a cuspy dark halo can disturb the central density 
profile and set up a constant density core and follow-up work with 
improved methods and resolution confirmed that result for their 
particular choice of bar parameters \citep{wei07b}.
\citet{sel08} has recently verified these results using
independent methods but has questioned the applicability of the results
of the dynamics of an idealized thin, rigid bar to real barred  
galaxies.  The model described here differs from these studies by examining a
{\em self-consistent model} of a bar forming from an instability in an
exponential disk within a cusped dark halo and so arguably represents a 
system closer to reality.  A detailed characterization of
the model here will allow us to compare our results to these other  
studies.

\begin{figure*}
\begin{center}
\epsscale{1.0}
\plotone{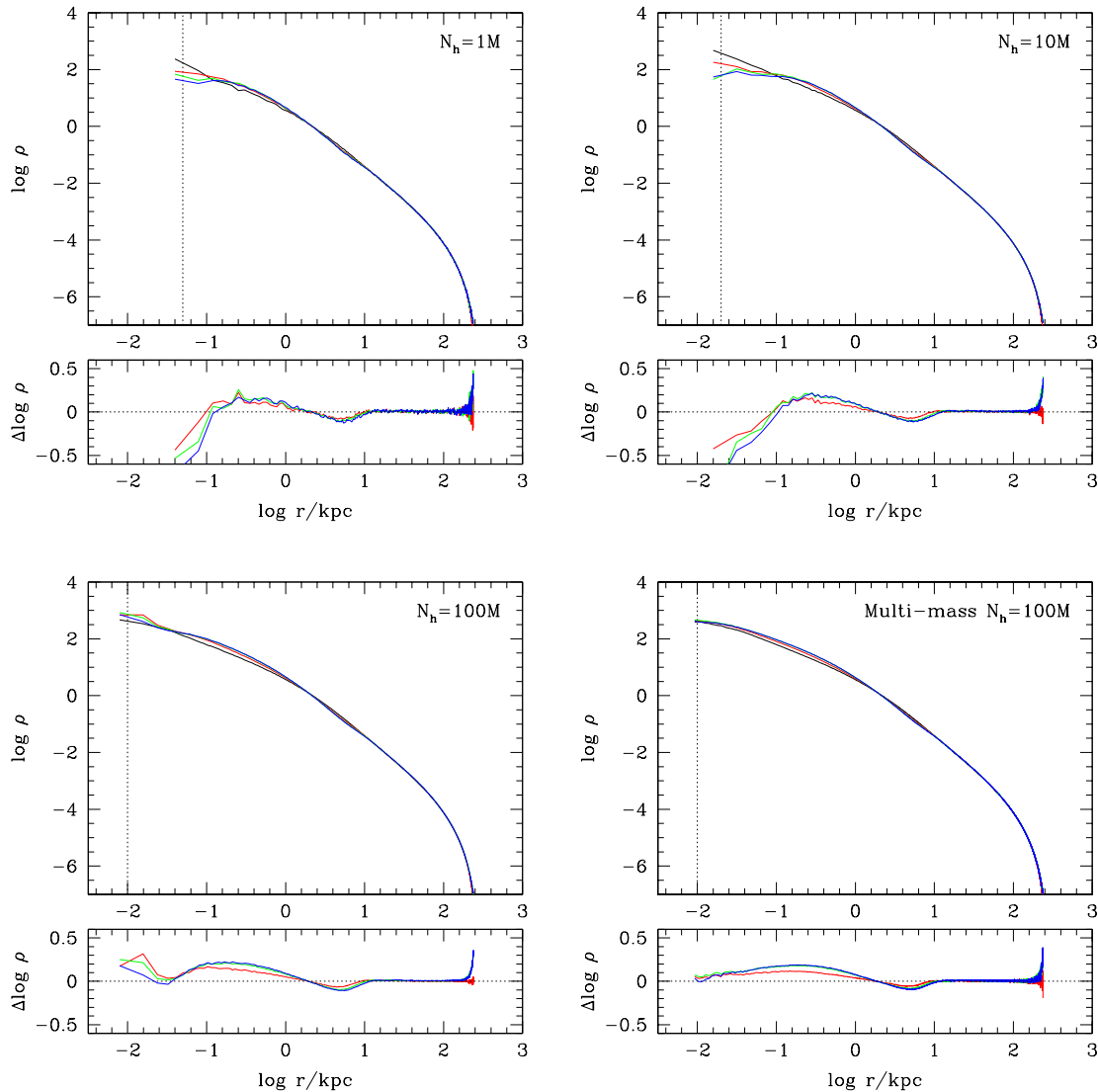}
\figcaption{
Evolution of the dark halo density profile for different mass resolutions 
at $t=0$, 2.4, 4.7 and 7.1~Gyr (black, red, green and blue lines).  The
dotted vertical line shows the value of the Plummer softening length for
each resolution: $\epsilon=50$,20,10 and 10 pc for $N_h=10^6, 10^7, 10^8$
and $10^8$ (multi-mass) halo models respectively.
For $r>10$ kpc, the density profile does not change
significantly.  In the range $1<r<10$ kpc, the density increases  
roughly $1.7\times$, showing adiabatic contraction in response to 
the  buckling instability and the formation of a centrally concentrated 
bulge-like bar within the disk.  The logarithmic slope of the density profile 
is $\alpha \sim -1$ to  within a few softening lengths from the center.  
A constant density core develops within the center with a core radius that 
depends on $N_h$ and  $\epsilon$ with typical values of $\sim 5\epsilon$.  
As $N_h$ increases and $\epsilon$ decreases, central density increases while 
the core radius declines.  The existence of a small core is consistent 
with relaxation due to softened gravity rather than forcing by the bar.
\label{fig-den-all}
}
\end{center}
\end{figure*}

\begin{figure}
\begin{center}
\plotone{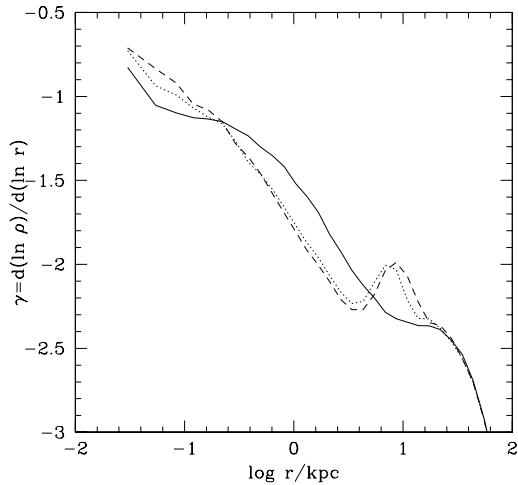}
\figcaption{
Change in the density power-law index profile $\gamma = d\log\rho/d \log r$
at $t=$0 Gyr (solid), 4.7 Gyr (dotted) and 7.1 Gyr (dashed).  The  
density profile maintains a cusped profile with $\gamma < -1$ down 
to $r=0.1$ kpc well within the the scale radius, $r_{-2} \approx 5$ kpc
at late times.  A constant density core does not develop in response 
to the bar and the halo maintains its cusp to the limit of gravitational
softening.
\label{fig-gamma}
}
\end{center}
\end{figure}

Figure \ref{fig-den-all} presents the evolution of the density profile
as a function of mass resolution.  The plots show the profiles at
changing times along with the differential change with respect to the
initial profile.  Gravitational softening introduces an artificial  
density core within $\sim 3$ Plummer softening lengths but beyond this  
radius the plots clearly show similar behavior in the density profiles.
A comparison of the final density profiles at different resolutions  
again shows similar behavior beyond the softening radius and convergence 
to a similar central behavior.  The central density profile
actually increases by $1.7\times$ while maintaining a central  
cusp (Fig. \ref{fig-gamma}).  The likely cause of this increase is 
the halo response to the  forming bar \citep{sel03,col06}.  
Once the bar buckles it forms a more concentrated mass distribution in 
the center of the disk and the halo responds by contracting adiabatically.
In the multi-mass halo with $N_h=10^8$, the density cusp is present down 
to $r \approx 100$ pc where gravitational softening effects start 
to influence the dynamics.  Within this radius, the halo is well-sampled by
more than 6000 particles and flattens out into a constant density core
dominated by softened gravity.

\citet{wei07b} (WK herein) have recently shown that massive bars can 
decrease the central density of DM halos and disrupt the cusp 
over a Hubble time in some cases at radii of about 20\% of the bar length.
Our bar has a length $a_b=12.5$ kpc so we should expect to see
distortions of the density profile at $r\approx 2$ kpc while in fact we see no
signs of a density core developing until softened gravity dominates at 
$r=0.1$~kpc in our highest
resolution case.  The results presented here seem to be in contradiction
so what's going on?  The reasons for disagreement can be understood 
by comparing the detailed 
properties  of the bars used in their models to our self-consistently evolved 
$N$-body bar.  The WK models are rigid, homogeneous ellipsoids of various 
masses, lengths and axis ratios rotating  within a live, isotropic
$N$-body halo.  Their fiducial model which strongly modifies the halo inner
profile has a bar length equal to the NFW halo scale radius $r_s$, i.e.,  
$a_b/r_s = 1.0$, a bar mass equal to half the halo mass within this 
radius $M_b/M_h = 0.5$ and an axis ratio  $a_1:a_2:a_3=10:2:1$.
Our halo is also NFW-like but not precisely a NFW model
due to modifications introduced in setting it up with an embedded  
disk and changes induced by bar formation.  A good proxy for $r_s$ in our  
models is the radius $r_{-2}$ where the density power-law slope
$\gamma=d\log\rho/d\log r \approx -2$ (For an NFW model $\gamma=-2$ at
$r=r_s$).  Figure \ref{fig-gamma} shows that $r_{-2} \approx 3$ kpc initially.  
At late times, the $\gamma$  profile develops a wiggle so that 
$\gamma=-2$ occurs at two different radii, but the average of these
two radii is $r_s\approx 5$~kpc.  The final bar length is about  $12$ kpc
(Fig. \ref{fig-contour}) and so $a_b/r_{-2}=2.4$.  The $N$-body
bar axis ratio measured above is $a_1:a_2:a_3 = 3.6:1.4:1$ considerably 
fatter than the fiducial model of WK.  Finally, the bar-to-halo mass ratio 
within the bar length is  $M_b/M_h=0.6$.  The main differences between 
the WK fiducial model and our model is that their halo is more extended 
and the bar is much thinner overall while the mass ratios are comparable.  
The WK models with thicker bars with $a_2/a_1>0.3$ are the closest ones to 
our $N$-body models and according to their Fig. 13 in \citet{wei07b}
cause no appreciable change in the density profile.  So we find no 
inconsistency with their most closely matching model.

While the thin, massive bars described by WK have strong effects on halo
profiles,  the thicker bars that develop through the recurrent buckling 
instabilities are more relevant to the evolution of real barred galaxies. 
Thin bars are subject to the dynamical buckling instability and thicken quickly.
Moreover, the strongest bars, i.e., those with $b/a\sim 0.2$ show a rapid  
decrease in the phase space available to regular orbits and hence an increase  
in the fraction of chaotic orbits in the bar \citep{mar04}.
While vertically thin {\it rigid} bars are immune to any instabilities, 
the DM particle orbits in the cusp can be destabilized
by the mere presence of a more massive analytical potential mixed  
with the live potential.  Vertically thinner bars, i.e., smaller $c/a$, 
will be more efficient in destabilizing the DM trajectories, by analogy 
with smaller $b/a$. In any case, the $c/a = 0.1$ thin bars used by WK cannot 
be justified over a Hubble time.  They are supported neither by 
observations or high-resolution  numerical simulations.

We conclude that bars that form self-consistently in $N$-body simulations 
from the instability of an exponential disk in NFW-like DM halos
do not destroy the density cusp and in fact can increase the
halo central density slightly.  Our mass resolution study shows a
clear convergence in behavior to higher resolution and the central
characteristics of dark halos are limited only by the particle  
softening and diminishing particle numbers.  We now explore the detailed 
orbital dynamics of the bars to understand angular momentum transport from 
the bar to the halo through low order resonances.

\begin{figure*}
\begin{center}
\plotone{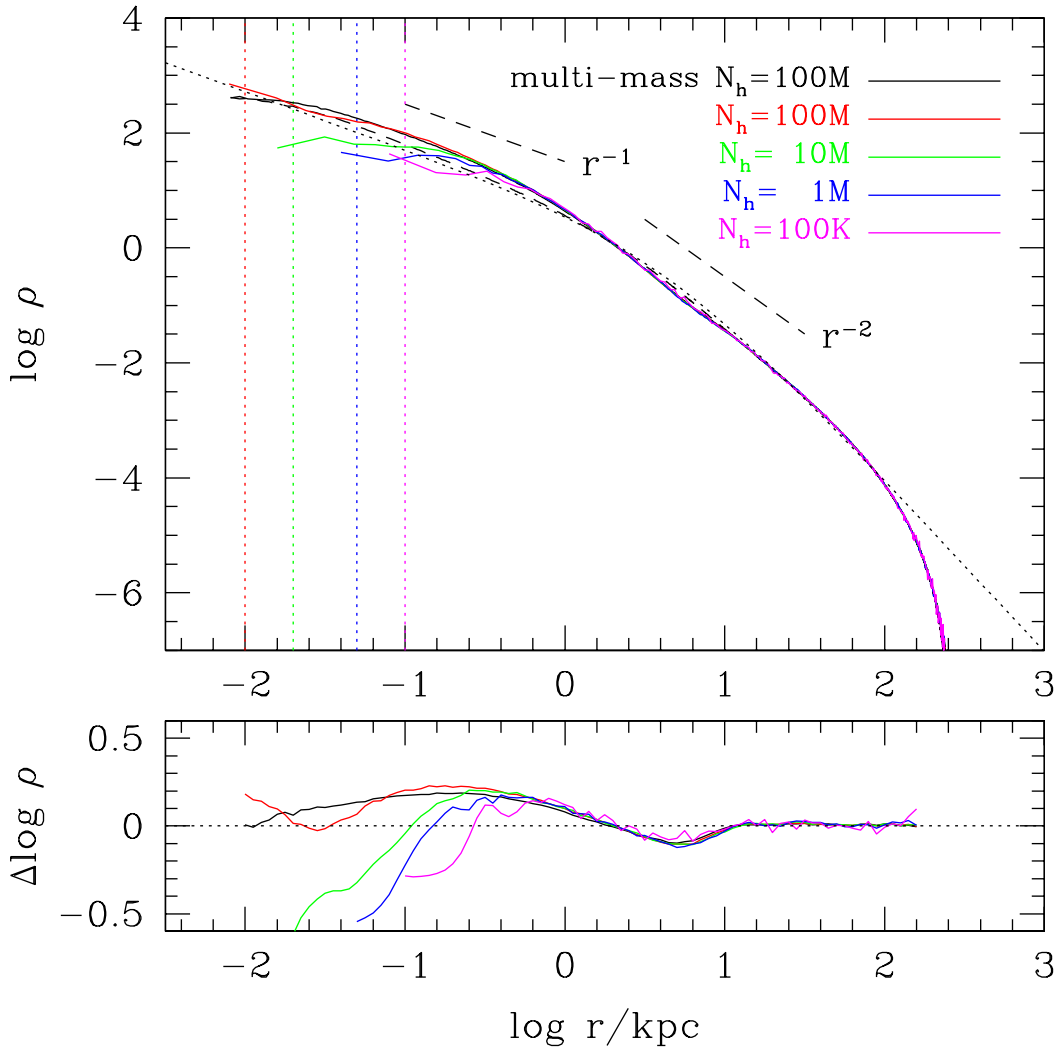}
\figcaption{
A comparison of density profiles at $t=7.1$~Gyr for different halo
particle numbers $N_h$.  We also show the initial density profile (dashed
line) and the best fit NFW model curve (dotted line) to the initial
profile over the range $0<r<100$~kpc.
The NFW parameters for the fit are $r_s=4.3$ kpc, $v_{max}=160$~km~s$^ 
{-1}$, where $v_{max}$ is the maximum circular velocity at
 $r=2.16 r_s=9.3$~kpc.  Note that this halo is more concentrated than 
the typical galactic dark matter halos in cosmological simulations. We use 
the NFW formula to characterize the profile and show that an $r^{-1}$ cusp  
extends to within at least 100 pc of the center.
The dotted vertical lines show the softening length $\epsilon$ used at
different resolutions.  As $N_h$ increases, the central density increases 
and the core radius decreases suggesting that the core behavior is due 
to mass resolution rather than forcing by the bar.
\label{fig-den-final}
}
\end{center}
\end{figure*}
\pagebreak

\section{Bar Orbital Dynamics}

Angular momentum is transferred from the bar to the halo through
low order orbital resonances \citep{lyn72,tre84,wei85}.
Following the convention of \citet{wei07a}, 
the condition for planar resonances is 
$l_1 \Omega_r + l_2 \Omega_\phi = m \Omega_b$ 
where $(l_1,l_2,m)$ are an integer triplet with radial and azimuthal
orbital frequencies $\Omega_r$ and $\Omega_\phi$ and 
bar pattern  speed $\Omega_b$ \citep{ath02,wei07a}. 
In the discussion below, we also use the parameters
$\Omega\equiv \Omega_\phi$ and $\kappa\equiv\Omega_r$ to refer to
the true orbital frequencies rather than the epicyclic approximations.
Bars are predominantly a $m=2$ disturbance so integer 
pairs $l_1:l_2$ with $m=2$ correspond to various
resonances with the more important ones being the inner and outer
Lindblad resonances (ILR $-1:2$ and OLR $1:2$) and the corotation 
resonance (COR $0:2$).  Other important resonances that may transfer
angular momentum occur with $l_2=-2,0$ including the direct radial 
resonance (DRR 1:0) discussed by WK.

We focus our analysis on resonances with $l_2=2$ that are responsible for
the bulk of angular momentum transfer.
A simple way of characterizing the low order resonances is with the
dimensionless frequency $\eta=(\Omega-\Omega_b)/\kappa$ \citep{ath02,mar06}.  
The half integer values of $\eta$ correspond to low order resonances with
$\eta=-1/2,0,1/2$ corresponding to the OLR, COR, and ILR respectively.
Most angular momentum is transferred to and from orbits
that satisfy this resonant condition.  As the bar loses angular momentum
and $\Omega_b$ declines, the population of halo particles in resonance
with the bar changes.  The potential of the halo also readjusts  
in response to the bar, so orbital frequencies can change as well.  
\citet{wei07a} have  argued that the resonances may only occur over a 
small fraction of the halo mass so that poorly resolved halos may not have 
sufficient numbers of particles to absorb angular momentum.  
Furthermore, noise in lower resolution simulations can cause particles 
to move in and out of resonance in a diffusive manner leading to an 
incorrect determination of angular momentum transfer.  They estimate that 
as many as $10^8$ halo particles are necessary to both populate resonances 
and suppress noise to converge on the correct behavior.  We examine 
these effects directly at different resolutions by studying 
the behavior of angular momentum transfer and orbital resonances using 
our models at the recommended mass resolution and see if 
the results do converge.

\subsection{Net Angular Momentum Transfer}

\begin{figure}
\begin{center}
\plotone{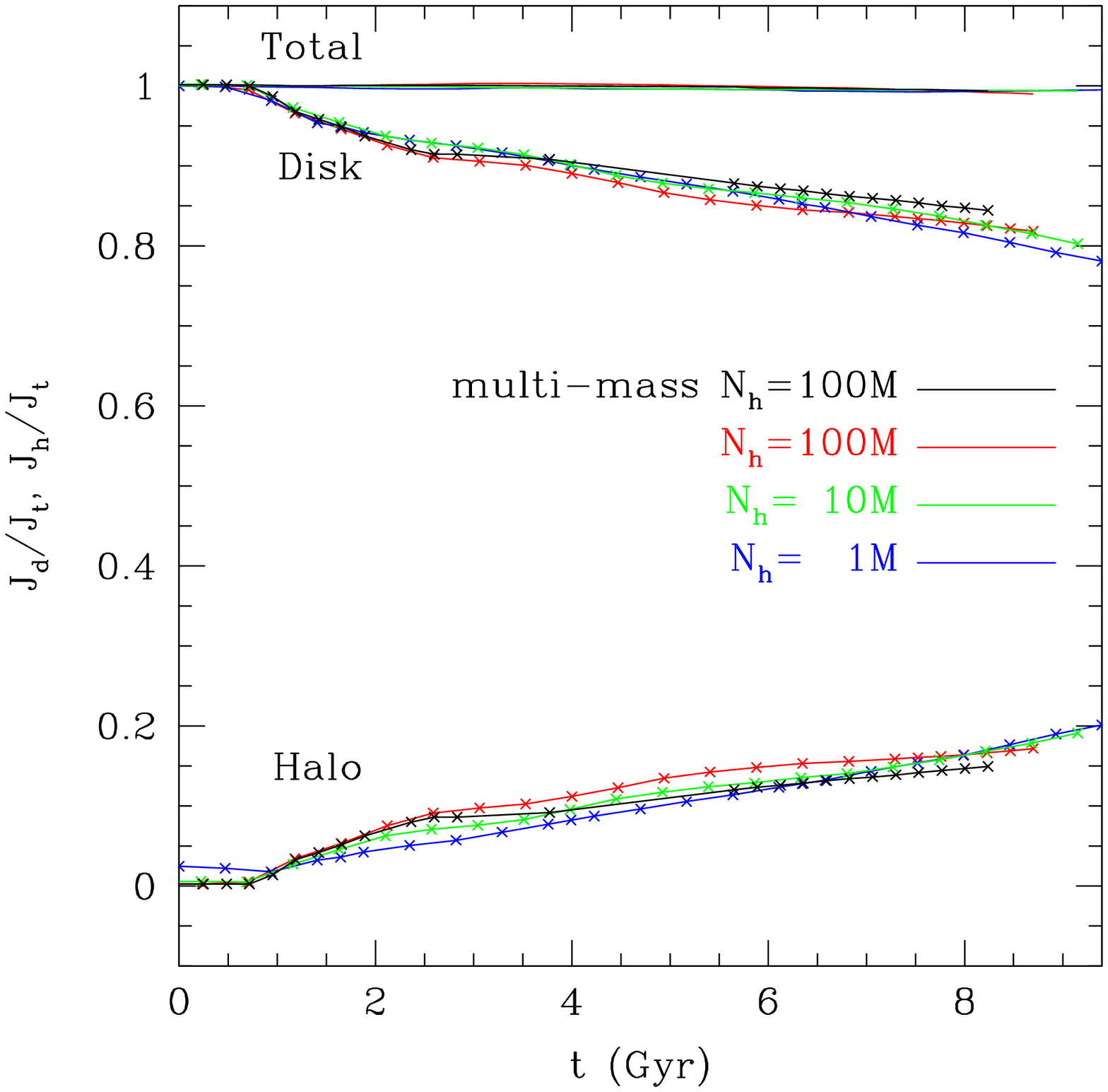}
\figcaption{
Evolution of the net angular momentum in the disk and halo at different
mass resolution.  Total angular momentum is conserved to within 1\%.
The evolution is similar for all resolutions plotted starting at
$N_h=10^6$.  The rate of angular momentum transfer is slightly smaller 
at later times for higher resolution simulations leading to about 
a 10\% difference in the total amount of transferred angular momentum in 
the multi-mass case suggesting a significant but small affect due to
resolution.
\label{fig-jvst}
}
\end{center}
\end{figure}

We first examine the net angular momentum transfer evolution as a function of
mass resolution (Fig. \ref{fig-jvst}).  We have again offset the  
times at different resolution so that they are synchronized with 
the time of maximum $|A_2|$.  The initial behavior is similar though there 
is no clear trend in behavior between resolutions from $t=2.4-7.2$ Gyr reflecting 
the variance from different random initial conditions.  At late times, however,
the rate of angular momentum transfer from the bar to the halo  
depends on resolution, with lower resolution simulations transferring $J$  
more quickly than the highest resolution case. From $t=7.2-9.4$ Gyr the rate 
of change $J$ is about two times larger for $N_h=1M$ than $N_h=100M$.  
This effect could be the result of noise broadening the resonant 
interaction though this interpretation is complicated by the variance 
in behavior due to different initial conditions.  In summary, there is 
a measurable difference in angular momentum transfer between high and low 
resolution with the  lowest resolution model transferring about 10\% 
more angular momentum.

\subsection{Halo Orbital Resonances}

We quantify the importance of low order resonances for angular momentum
transfer in our models using a modified version of the orbital  
spectral analysis method \citep{bin82} in a frozen rotating potential
as described by \citet{ath02} and \citet{mar06}.
We determine the principal orbital frequencies $\kappa$ and $\Omega_\phi$
for a set of $N_{Rh}$ randomly chosen particles
in the halo with $N_{Rh}\sim 10^6$ for simulations with $N_h \ge 10^6$
and $N_{Rh}=N_h$ for smaller simulations.  We then compute the potential and
force field on a grid with variable spacing to be used for interpolating forces
for test particle integrations.  Orbital frequencies are 
determined from test particle integration of particles orbits in 
the frozen potential in a rotating frame at the bar's pattern speed for 
the time of a given snapshot.  The orbits are integrated for about 
50 bar rotations, starting at three representative times --- 
$t = 2.4,  4.7$ and 7.1 Gyr for the $N_h=10^8$ single and multi mass simulations.  
Appropriate time offsets are applied as discussed above to lower 
$N_h$ simulations to synchronize the time of maximum $|A_2|$.  
Each orbit was sampled with at least 200 constant timesteps per azimuthal period, 
and overall by 10K timesteps. The use of constant timesteps simplifies 
Fourier decomposition of the orbit time series.  Most decompositions lead 
to a line spectrum allowing easy identification of frequencies 
$\Omega$ and $\kappa$ though occasionally the spectrum is 
more complex and no frequencies can be uniquely identified.  

\begin{figure}
\epsscale{1.1}
\begin{center}
\plotone{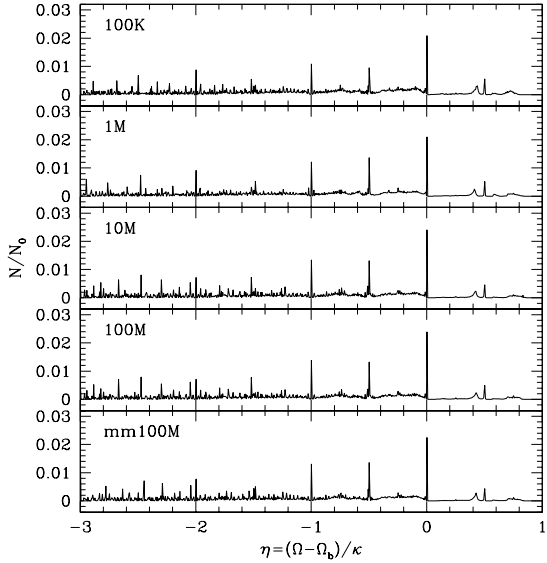}
\figcaption{
Distribution of DM halo particles as a function of the dimensionless frequency
$\eta$.  Resonant spikes at the half integer values of $\eta$ correspond to
low order resonances.  The bin width is $\delta \eta=0.005$. 
The distributions are similar as a function of mass resolution.
\label{fig-jspec}
}
\end{center}
\end{figure}

We present the results of the spectral analysis for DM halo orbits in 
Fig.~\ref{fig-jspec}
for various resolutions at $t = 7.1$ Gyr for the 100M particle runs.  Again, 
we account for the time offsets discussed above for the lower resolution 
runs for a fair comparison.  The  particles are 
binned in frequency $\eta$ with a bin width $\Delta\eta = 0.005$.
Figure~\ref{fig-jspec} shows the distribution of the particle number
fraction (or mass fraction in the multi-mass model case) as a function of
the dimensionless frequency.  The main resonances -- ILR, COR, OLR - are
present along with higher order ones with COR being the most populated
resonance.  The relative height of the peaks begins to converge when
$N>10^6$ and the behavior is quite similar.  The peak bins contain a few
percent of the total particle numbers or $\sim 10^6$ particles in the
largest case and so provide good coverage of the resonance for angular
momentum transfer.  We can define the amount of mass in
``resonance" as the sum over particles with dimensionless frequencies 
in the range $\delta \eta \pm 0.05$ at half integer values of $\eta$.
When measured this way about 7\% of the total halo mass is in resonance
instantaneously at late times when the bar has reached quasi-equilibrium and
is slowing down.   \citet{tre84} speculated that orbits may have become
trapped in resonance if the bar slowed down gradually and we can check whether
this trapping is significant.
A comparison of the particles in resonant peaks at $t=4.7$ Gyr
with those at $t=7.1$ Gyr shows that only a small fraction migrate between 
resonances as the system evolves.  Of the 7\% of the total mass in resonance at $t=4.7$ Gyr only 
1.5\% are still in resonances at $t=7.1$ Gyr with most particles moving
out of resonance.  As the bar is braking,  
new orbits are brought into resonance while orbits 
that have acquired angular momentum move out of resonance.  In this sense, 
the resonance is broad and a significant fraction of halo orbits participate 
in angular momentum exchange with the bar.

\begin{figure}
\epsscale{1.1}
\begin{center}
\plotone{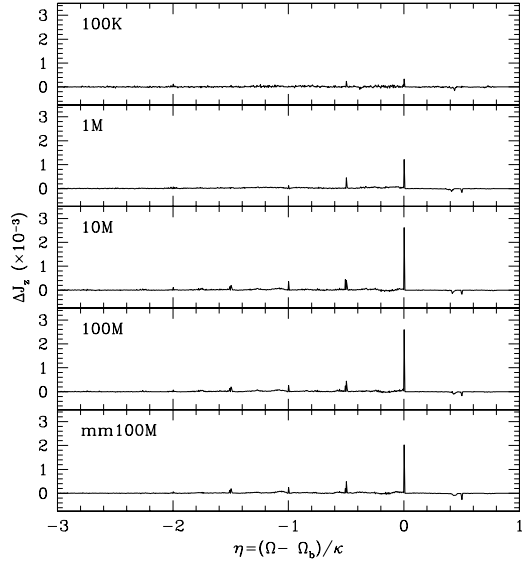}
\figcaption{
Net change in the DM halo particle angular momentum 
between $t=4.7$ and 7.1~Gyr for particles binned as a function of the 
dimensionless frequency $\eta$ measured at
$t=4.7$ Gyr.  The majority of angular momentum is gained
through the CR resonance at $\eta=0$ though some angular momentum is lost
at the ILR at $\eta=0.5$.  The peaks are sharper at higher resolution.
\label{fig-dj}
}
\end{center}
\end{figure}

\begin{figure}
\epsscale{1.1}
\begin{center}
\plotone{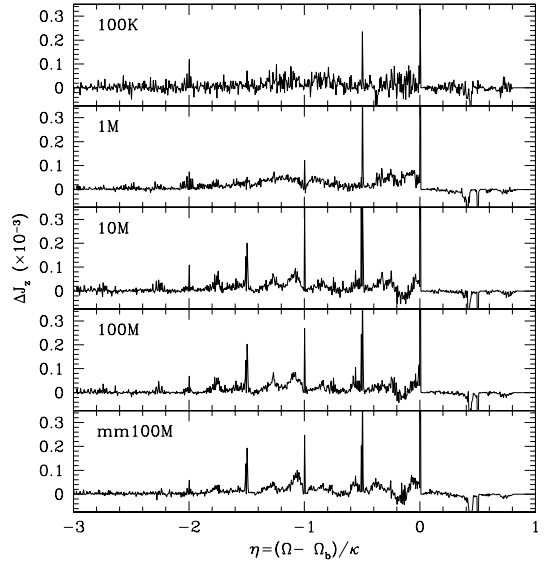}
\figcaption{
Same as Fig.~\ref{fig-dj} with the vertical scale expanded by $10\times$.
The detailed distributions of the change in angular momentum are similar
between the peaks at higher resolution.  
\label{fig-dj1}
}
\end{center}
\end{figure}

Figures~\ref{fig-dj} and \ref{fig-dj1} show the resonant transfer of
angular momentum between the two snapshots at $t=4.7$ Gyr and $t=7.1$ Gyr.  
We plot the distribution of the change in z-angular momentum 
$\Delta J_z$ versus $\eta$ measured for the particles at
$t=4.7$ Gyr.  Most angular momentum is absorbed in the halo at the COR and
OLR with smaller amounts absorbed at higher order resonances.  However,
some $J_z$ is emitted and lost from the ILR in accord with fundamental ideas of
angular momentum transport in stellar systems \citep{lyn72}.  The
distributions when viewed with an expanded vertical scale show nice
convergence in detailed behavior at higher resolution (Fig.~\ref{fig-dj1}).  
For $N_h>10^7$, $\approx 50$\% of the total transferred angular 
momentum is is in the resonant peaks (within $\delta \eta = \pm 0.05$
while for $N_h\le 10^6$ we find less than 30\% in the peaks with this same
definition.  The lower resolution simulations are clearly more susceptible
to diffusion.  Nevertheless, despite these differences the total angular
momentum transferred is similar for $N\ge 10^6$ suggesting that the
diffusive process that broadens resonances is not a serious problem for the
global evolution of the system.

\pagebreak

\subsection{Resonances in Phase Space}

\begin{figure}
\epsscale{1.2}
\begin{center}
\plotone{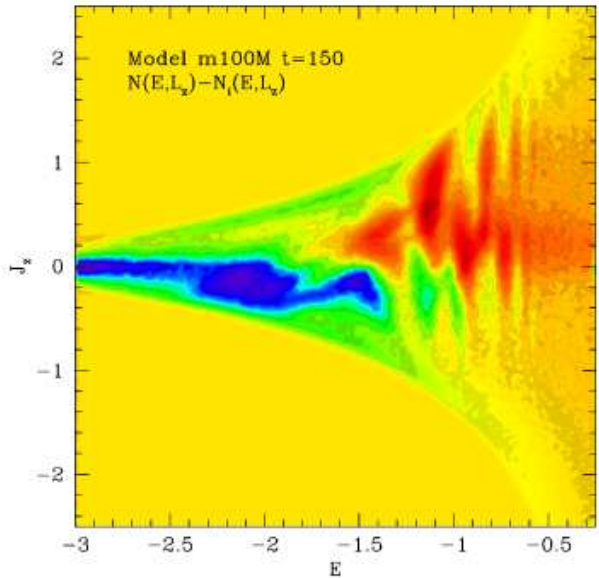}
\figcaption{
Change in particle number density in $(E,J_z)$ space between $t=0$ and
$t=150$ (7.0 Gyr) for the $N_h=10^8$ single mass model.  The resonant 
regions show up clearly as peaks (red regions) in phase space in the 
left panel.  The blue-black region is a valley where a halo bar rotating
along with the disk bar and so de-populated the negative $J_z$ of phase
space at the ILR. See Video 4 to view the time evolution of the particle
phase-space density.
\label{fig-phase}
}
\end{center}
\end{figure}

\begin{figure}
\epsscale{1.2}
\begin{center}
\plotone{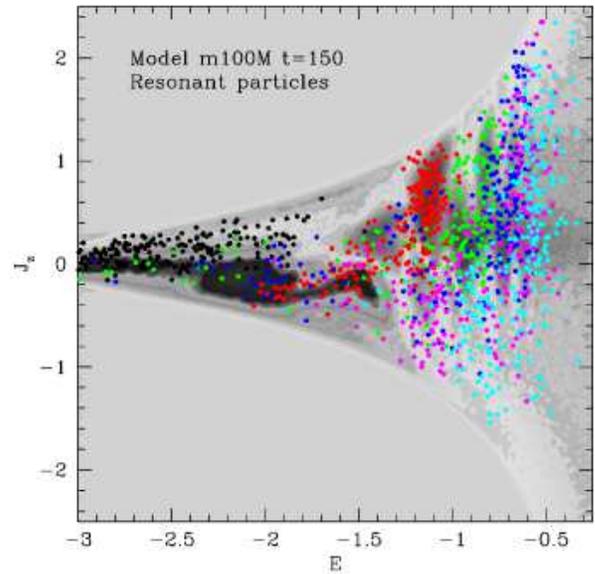}
\figcaption{
On the phase number density map, we overplot the $(E,J_z)$ coordinates of
a subset of particles located at discrete resonances at $t=150$ 
within $\delta\eta=\pm 0.05$ (black-ILR-$\eta=0.5$, red-COR-$\eta=0.0$, 
green-OLR-$\eta=-0.5$, blue-$\eta=-1.0$, magenta-$\eta=-1.5$, 
and cyan-$\eta=-2.0$.  The resonant particles lay directly on top of the 
peaks and so identify the specific resonant regions in phase space.
\label{fig-resonant}
}
\end{center}
\end{figure}

Finally, we examine the change in halo phase space density by computing the
particle number density in $(E,J_z)$ space and computing the difference
between $t=0$ and $t=150$ in model m100M in a similar way to 
\citet{hol05}.  In this way,
we clearly see the resonant regions visible as discrete islands of particle 
overdensity in $(E,J_z)$ space (Fig.~\ref{fig-phase}).  We can also overplot 
the values of $(E,J_z)$ for the particles found in the resonant spikes 
in the analysis at the final time $t=150$ to see where they lie in 
phase space.  Figure~\ref{fig-resonant} clearly shows that the peaks
in phase space density are directly related to the discrete resonances 
extracted from our spectral analysis.  An accompanying animation to
Fig~\ref{fig-phase} presents the time evolution of the differential number 
density in phase space and reveals how the resonant islands move through a 
large fraction of the halo mass.  By counting particles in resonant peaks 
at different times we estimate that roughly 20-30\% of the halo particles 
are in resonance with the bar at some time in their history.  Since such a
large fraction of particles are involved in angular momentum transfer then
even lower resolution simulations can do a reasonable job of following the
evolution of the bar.

\section{Conclusions}

We have carried out a comprehensive set of experiments to explore the
evolution of a self-consistent bar in a galactic model with an exponential
disk and cuspy DM halo using resolutions
with $10^{4-8}$ DM particles and a single experiment using a
multi-mass method with an effective resolution of $10^{10}$. 
Our highest resolution exceeds by far the level prescribed by \citet{wei07a}
necessary to achieve convergent behavior in bar galaxy dynamics.
We have applied various diagnostics of bar evolution as a function of  mass
resolution including bar growth, pattern speed evolution, halo density cusp
evolution and the resonant transfer of angular momentum from the bar to the
DM halo.  In almost all cases, the general behavior is similar at most but
the lowest resolutions with the convergence occurring around $10^{6-7}$, 
depending on the phenomenon.   
Sellwood (2008) has also explored similar effects in a mass resolution study
with rigid bars in cuspy spherical halos with $\sim 10^8$ particles and
come to similar conclusions about minimal resolution requirements.
Notably, in this model the
density cusp is not destroyed by the formation of the bar in apparent
contradiction to the results of WK.  Our best explanation is that the thick bar
that form in our self-consistent models has a weaker affect than the rigid
thin bars in the work of WK and we question the applicability of these thin
bar models over a Hubble time in light of the buckling instability. 

The strongest argument for convergence comes from the spectral analysis of
orbits in the rotating barred potential at different resolutions 
that shows in detail similar distributions as a function of the
dimensionless frequency $\eta$ both in mass fractions and angular momentum
transferred between different times.  Analysis of the change in
phase space density show that resonant islands sweep through the phase
space as the bar loses angular momentum leading to effectively broader
resonances with as much as 20-30\% of the halo mass absorbing angular
momentum from the bar.

Future studies should examine the bar instability self-consistently using 
the same initial conditions with different $N$-body methods to resolve 
current inconsistent results on the cusp/core evolution of DM halos as well
as explore detailed behavior in phase space.  The model snapshots and initial
conditions from this study are freely available to researchers in the
area who wish to verify our results against their own codes and
methods.

\acknowledgements{We acknowledge useful discussions with Jerry Sellwood,
Simon White, James Binney, Linda Sparke and Larry Widrow.  We also thank
the referee for useful comments.  This work 
was supported, in part, by the NSERC 
of Canada and the Canadian Foundation for Innovation.   
I.S. acknowledges JILA Visiting Fellowship and partial 
support from NASA/ATP/LTSA, NSF and  
the STScI. I.B. acknowledges financial support from the Volkswagen
Foundation (Ref: I/80 041-043). Supercomputing 
was provided by SHARCNET  facilities at McMaster University and the 
University of Waterloo as well as  facilities at CITA. }

\bibliographystyle{apj.bst}
\bibliography{refs}

\end{document}